\documentclass[superscriptaddress,showpacs,twocolumn,aps,pre,10pt]{revtex4-1}

\usepackage{graphicx,dcolumn,bm,bbm,amsmath,amssymb,color}
\usepackage[squaren]{SIunits}
\usepackage[english]{babel}

\begin{document}

\title{Brownian motion of a circle swimmer in a harmonic trap}

\author{Soudeh Jahanshahi}
\affiliation{Institut f\"ur Theoretische Physik II: Weiche Materie, Heinrich-Heine-Universit\"at D\"usseldorf, D-40225 D\"usseldorf, Germany}

\author{Hartmut L{\"o}wen}
\affiliation{Institut f\"ur Theoretische Physik II: Weiche Materie, Heinrich-Heine-Universit\"at D\"usseldorf, D-40225 D\"usseldorf, Germany}

\author{\surname{Borge} ten Hagen}
\email{b.tenhagen@utwente.nl}

\affiliation{Physics of Fluids Group, Faculty of Science and Technology, University of Twente, 7500 AE Enschede, The Netherlands}

\begin{abstract}
We study the dynamics of a Brownian circle swimmer with a time-dependent
self-propulsion velocity in an external temporally varying harmonic potential.
For several situations, the noise-free swimming paths, the noise-averaged mean trajectories, and the mean square displacements are calculated analytically or by computer simulation. Based on our results, we discuss optimal swimming strategies in order to explore a maximum spatial range around the trap center. In particular, we find a resonance situation for the maximum escape distance as a function of the various frequencies in the system. Moreover, the influence of the Brownian noise is analyzed by comparing noise-free trajectories at zero temperature with the corresponding noise-averaged trajectories at finite temperature. The latter reveal various complex self-similar spiral or rosette-like patterns. Our predictions can be tested in experiments on artificial and biological microswimmers under dynamical external confinement.
\end{abstract}

\pacs{82.70.Dd, 05.40.Jc}

\maketitle

\section{Introduction}
\label{introduction}

Trapping of classical particles in a dynamically changing environment
occurs in many situations ranging from colloids in optical tweezers
\cite{Grier1997,Deng2007} to the motion of tracers in a fluctuating
host matrix \cite{Lehmann2012}. It is also of fundamental
importance for understanding Brownian motion in a time-dependent external
potential \cite{Haenggi1990}.
The simplest nontrivial setup is a Brownian particle in a harmonic
external trapping potential with a time-dependent prefactor. This
model has been extensively studied as the Brownian parametric oscillator
\cite{Zerbe1994, Brouard2001} and as a basic model for multiplicative noise \cite{Barzykin1998, Berdichevsky1999,
Gitterman2005, Calisto2006, Mankin2008}, for stochastic resonance \cite{Gammaitoni1998, Barzykin1997}, and for fluctuation
squeezing \cite{Tashiro2007}. More recently, it has become popular to use this
model to discuss the efficiency of nonequilibrium work production
\cite{Kwon2013,Ryabov2013} and stochastic heat engines \cite{Schmiedl2008,Holubec2014}.
In the limit of completely overdamped Brownian motion, the stochastic
equations of motion can be solved analytically, resulting in a time-dependent
Gaussian process for the particle displacements. For special forms
of the time dependence, the analytical solution was further analyzed,
and a giant breathing effect \cite{Lowen2009}
was put forward for a potential that periodically flips between a
stable and an unstable situation. Moreover, an unusual scaling in the mean square displacement
of the particle was obtained when the harmonic confinement fades away
algebraically in time \cite{Chakraborty2012}.

In the past few years, microswimmers have increasingly been in the focus of research \cite{Romanczuk2012,Elgeti2015,Zottl2016,Bechinger2016}. Inspired by biological systems \cite{Berg1972,Goldstein2015}, several different propulsion strategies have been introduced to realize artificial microswimmers \cite{Paxton2004,Howse2007,Jiang2010,Ebbens2010SM,Volpe2011,Theurkauff2012,Buttinoni2013,Palacci2013Science}. These particles
self-propel and dissipate energy while they move and are therefore genuinely in nonequilibrium \cite{Marchetti2013,Menzel2015,Solon2015,Weber2011,Debnath2016}. 

One of the simplest basic models is to describe microswimmers as active
Brownian particles with an internal effective self-propulsion force \cite{tenHagen2015}
leading to a velocity $v_0$ that is attached
to the body-frame of the particle. The direction of propulsion fluctuates
according to orientational Brownian motion as characterized by a short-time
rotational diffusion constant $D_r$. In the two-dimensional bulk,
it has been shown \cite{Howse2007} that for long times the dynamics of these active Brownian particles
is a persistent random walk where the persistence length is $v_0/D_r$
such that the long-time diffusion coefficient is enhanced by the term $v_0^2/(2 D_r)$.
Higher displacement moments and the non-Gaussian behavior of the dynamical process
at finite time have also been calculated subsequently \cite{tenHagen2011}. The basic model of an active Brownian particle
can readily be generalized to situations where the self-propulsion velocity
$v_0(t)$ is time-dependent \cite{Babel2014} as it is the case for the run-and-tumble motion of many bacteria, for example. 

Several recent analytical and numerical studies \cite{tenHagen2011,Stark2012,Volpe2013,Szamel2014,Nourhani2015,Ribeiro2016}
consider the motion of a self-propelled particle in an external harmonic
trap. As a result, it was found that self-propulsion induces an increased
delocalization in the trap, which, however, cannot uniquely be described by an effective equilibrium temperature \cite{Tailleur2009,Szamel2014,BenIsaac2015}. In all of these previous
works on \textit{linear} swimmers, a \textit{static} harmonic
confinement and a \textit{time-independent}
self-propulsion velocity were assumed. In this paper, we generalize the situation to \textit{circle swimmers} \cite{vanTeeffelen2008,vanTeeffelen2009,Reichhardt2013separation,Kummel2013,Yang2014Y,Ao2015,Schirmacher2015,Geiseler2016PRE}, 
which perform an active rotational motion in addition to their translational self-propulsion, and consider both a \textit{time-dependent} harmonic trap and a \textit{time-dependent} self-propulsion
velocity. 
Our motivation to do so is threefold: First, we expect new physics
due to a constructive competition between internal swimming and external
switching degrees of freedom, which needs to be explored and analyzed.
Second, we obtain analytical results and any analytical solution in nonequilibrium physics is interesting
in itself since it provides an ideal building block for a minimalist
model for more complex systems. Third, self-propelled particles in a time-dependent external potential are relevant in various experimental situations. Examples discussed in this manuscript include self-diffusiophoretic microswimmers \cite{Volpe2011,Buttinoni2012,Kummel2013}, self-propelled particles in acoustic potentials \cite{Takatori2016}, and in a somewhat wider context also active granular hoppers \cite{Deseigne2010,Briand2016}. Moreover, real bacteria with
a characteristic run-and-tumble motion \cite{Berg1972} can be exposed to geometric and other confinements \cite{Galajda2007,Kaiser2014PRL,Vladescu2014}.
Thus, our model can be realized in experiments on artificial microswimmers or swimming microorganisms.  

In our study, we consider both the limit of zero noise and the situation with full Brownian motion. In the first case, subsequent to an initial regime which stems from the external potential, periodic trajectories are found. Their period is determined by an interplay between the frequencies of the circle swimming, the time-dependent self-propulsion, and the oscillating external potential. If thermal fluctuations are included, the mean trajectories are shown to be self-similar curves collapsing into the trap center. With regard to the mean square displacement, we provide a general analytical result which is explicitly solved for an active Brownian particle with temporally varying self-propulsion in a constant external potential. Furthermore, we discuss optimal swimming strategies for the particle in order to explore the largest spatial range inside the harmonic trap. In this context, a resonance situation is found for the maximum reachable distance from the trap center.

This paper is organized as follows: In Sec.\ \ref{model}, the basic model for a Brownian circle swimmer in a harmonic external potential is introduced. Subsequently, the influence of two types of time dependencies in the system is investigated. While the focus is on a time-dependent self-propulsion velocity in Sec.\ \ref{propulsion}, Sec.\ \ref{potential} is devoted to a temporally varying potential strength. In all cases, we first present results for vanishing noise before considering the general situation with Brownian motion. Section \ref{experiment} contains a discussion of various experimental realization possibilities for our model. Finally, conclusions and an outlook are given in Sec.\ \ref{conclusions}.

\section{The basic model}
\label{model}

\begin{figure}[tb]
\centering
\includegraphics[width = \columnwidth]{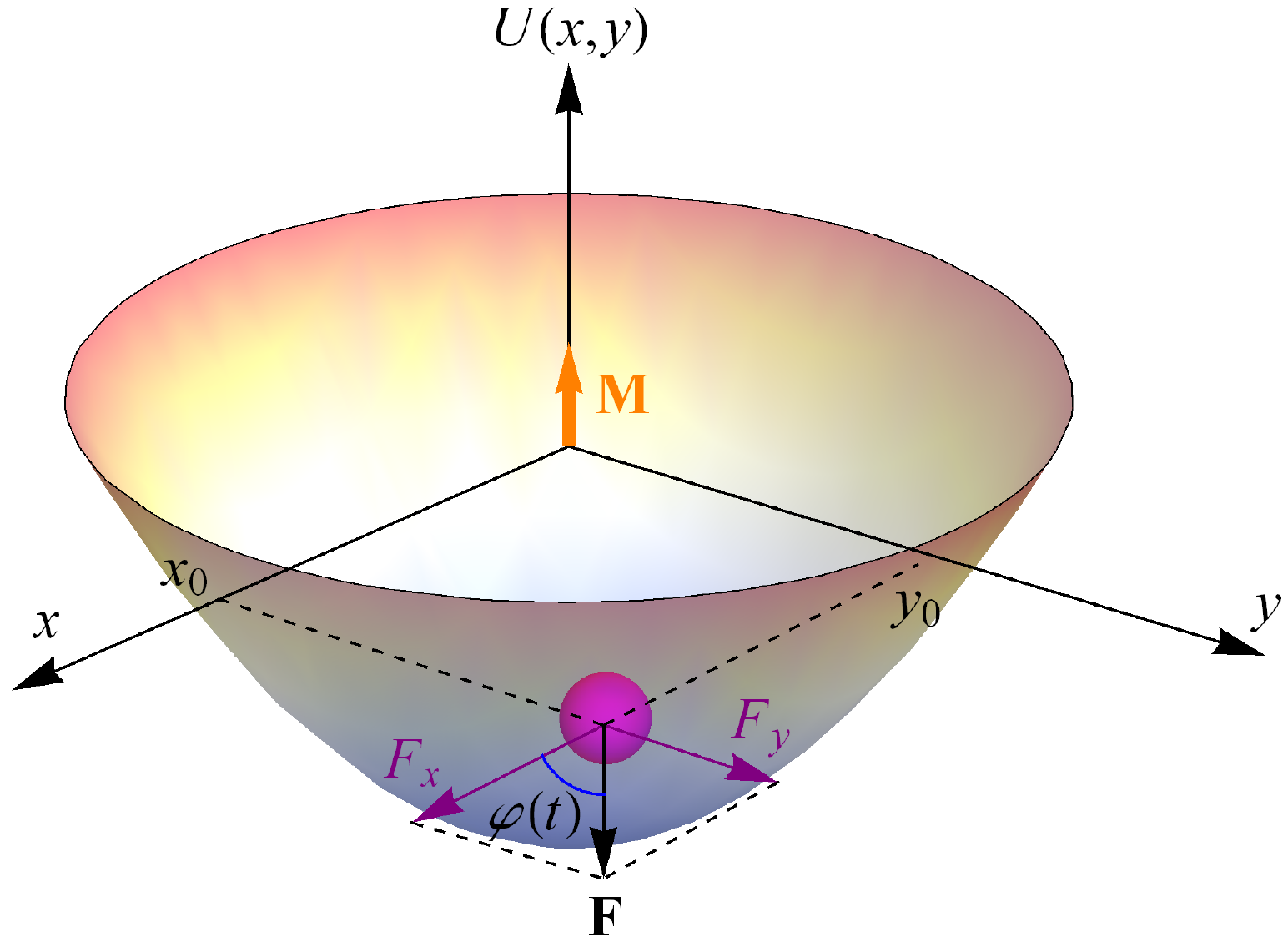}
\caption{\label{1g_1}(Color online) Schematic view of a spherical circle swimmer in a symmetric harmonic spatial trap $U(x,y)$. The self-propulsion is characterized by the effective force $\mathbf{F}=\left(F_x,F_y\right)$ and the effective torque $\mathbf{M}=M\hat{\mathbf{e}}_{z}$. As the rotational motion is restricted to the $xy$ plane, the particle orientation is fully determined by the angle $\varphi$.}
\end{figure}

To investigate the Brownian motion of a single circle swimmer in a harmonic trap in two spatial dimensions, we use corresponding overdamped Langevin equations. The external potential, which is assumed to be symmetric and centered on the origin, is given by
\begin{equation}
\label{eq:pot}
U\left(x,y\right)=\frac{\lambda_{0}}{2}\left(x^{2}+y^{2}\right)
\end{equation} 
with the potential strength $\lambda_0$. 
The self-propulsion is modeled by an effective force $\mathbf{F}=F_0 \hat{\mathbf{u}}$ \cite{tenHagen2015} along the particle orientation $\hat{\mathbf{u}}$ and an effective constant torque $\mathbf{M}$. In our model, we assume that both the translational and the rotational motion are restricted to two dimensions. In experiments, such a situation is often caused by hydrodynamic effects between the self-propelled particles and the substrate \cite{Das2015,Simmchen2016}. These effects lead to a configuration where the particle orientations are always almost parallel to the substrate. Hence, only a single angle $\varphi$ is needed in the theoretical description, and the particle orientation can be written as $\hat{\mathbf{u}}=(\cos\varphi,\sin\varphi)$. Furthermore, the torque $\mathbf{M}$ acts always along the perpendicular $z$ direction, i.e.\ $\mathbf{M}=M\hat{\mathbf{e}}_{z}$ (see Fig.\ \ref{1g_1}). 

In the overdamped limit, the translational motion of a
particle confined to the two-dimensional $xy$ plane is described by the Langevin equations
\begin{equation}
\frac{d}{dt}x\left(t\right)=\beta D\left[F_{0}\cos\left(\varphi \left(t\right)\right)-\lambda_{0} x\left(t\right)\right]+\sqrt{2D}\,\xi_{x}\left(t\right)
\label{Langevinx}
\end{equation}
and
\begin{equation}
\frac{d}{dt}y\left(t\right)=\beta D\left[F_{0}\sin\left(\varphi \left(t\right)\right)-\lambda_{0}y\left(t\right)\right]+\sqrt{2D}\,\xi_{y}\left(t\right)
\label{Langeviny}
\end{equation}
for the center-of-mass position $\mathbf{r}(t)=(x(t),y(t))$ of the particle.
$\xi_{x}(t)$ and $\xi_{y}(t)$ are independent Gaussian white noise random terms, $D$ denotes the translational short-time diffusion constant, and  $\beta=1/(k_B T)$ is the inverse effective thermal energy.
The terms $-\lambda_{0} x\left(t\right)$ and $-\lambda_{0} y\left(t\right)$ take the effect of the external potential [see Eq.\ \eqref{eq:pot}] into account. The rotational motion of the particle is governed by a single Langevin equation for the angle $\varphi\left(t\right)$,  
\begin{equation}
\frac{d}{dt}\varphi\left(t\right)=\sqrt{2D_{r}}\,\xi_{\varphi}\left(t\right)+\omega,
\label{Langevinphi}
\end{equation}
in which $\xi_{\varphi}\left(t\right)$ represents a zero-mean Gaussian
white noise random torque and $D_{r}$ is the
rotational short-time diffusion constant. For spherical
particles with radius $R$, $D$ and $D_{r}$ fulfill $D/D_{r}=4R^{2}/3$ \footnote{This ratio holds for bulk situations. For particles moving near walls or system boundaries, the diffusion coefficients usually have to be rescaled appropriately \cite{Goldman1967,Kummel2013}}. The angular frequency $\omega=M\beta D_{r}$, which is determined by the effective torque $M$ and leads to the chiral motion, may originate from particle imperfections \cite{Takagi2013} or from external fields \cite{Baraban2012,Rupprecht2016}, for example. Furthermore, circle swimming can be due to spontaneous symmetry breaking as has recently been observed for self-propelling nematic liquid crystal droplets \cite{Kruger2016}. When proceeding from spherical particles as considered here to more complex particle shapes, an active rotational motion usually follows directly from the shape asymmetry in combination with the detailed self-propulsion mechanism \cite{Kummel2013,Kuemmel2014}. Then, the hydrodynamic coupling between the translational and the rotational self-propelled motion plays an important role. This is manifested by the nonzero off-diagonal elements in the corresponding grand resistance matrix \cite{Brenner1967,Kraft2013}. In our theoretical model in the present manuscript, we do not focus explicitly on the specific origin of the active rotational motion but just include a constant angular velocity $\omega$, which is referred to as \textit{circling frequency} in the following.

The effect of random kicks of the solvent molecules is taken into account by means of the Gaussian white noise terms $\xi_{x}(t)$, $\xi_{y}(t)$, and $\xi_{\varphi}(t)$ in Eqs.\ \eqref{Langevinx}--\eqref{Langevinphi}. These terms are  characterized by $\langle \xi_{i}(t)\rangle  =0$ and 
$\langle \xi_{i}(t)\xi_{j}(t^{'})\rangle =\delta_{ij}\delta(t-t^{'})$ with $i,j \in \{x,y,\varphi\}$. To study the statistical behavior of the system, noise-averaged quantities, such as the mean trajectories and the mean square displacement, are calculated from the Langevin equations. 

According to Eq.\ \eqref{Langevinphi}, the angular coordinate $\varphi(t)$ is a linear combination of Gaussian variables $\xi_{\varphi}(t)$. Thus, the corresponding probability distribution function $P(\varphi,t)$ is Gaussian as well and is obtained by just evaluating the mean $\langle \varphi(t) \rangle = \varphi_0 +\omega t$, where $\varphi_0$ is the initial angle at $t=0$, and the variance $\langle (\varphi(t)-\langle \varphi(t)\rangle)^2\rangle = 2D_rt$ from Eq.\ \eqref{Langevinphi}. The result 
\begin{equation}
P\left(\varphi,t\right)=\frac{1}{\sqrt{4\pi D_{r}t}}\exp\left[-\frac{\left(\varphi -\varphi_{0} -\omega t\right)^{2}}{4D_{r}t}\right]
\end{equation}
allows for an analytical calculation of the noise average of any angular terms occurring in the Langevin equations for a self-propelled particle on a substrate \cite{tenHagen2009}. 

Most of the results in this manuscript are provided in a dimensionless form. Therefore, we define the dimensionless quantities $\lambda'_{0}=\beta\lambda_{0} D/D_{r}=4\beta\lambda_{0}R^2/3$ for the strength of the external potential, $\omega'=\omega/D_{r}$ for the circling frequency, and $F'_{0}=\beta\left(D/D_{r}\right)\left(F_{0}/R\right)=4\beta F_0 R/3$ for the effective self-propulsion force of a spherical particle. 
For reasons of presentation, we omit the prime symbols in the following. When exceptionally an equation is given in a nondimensionless form, this is explicitly mentioned.

\subsection{Results for vanishing noise}

\begin{figure*}[tb]
\centering
\includegraphics[width = 0.9\linewidth]{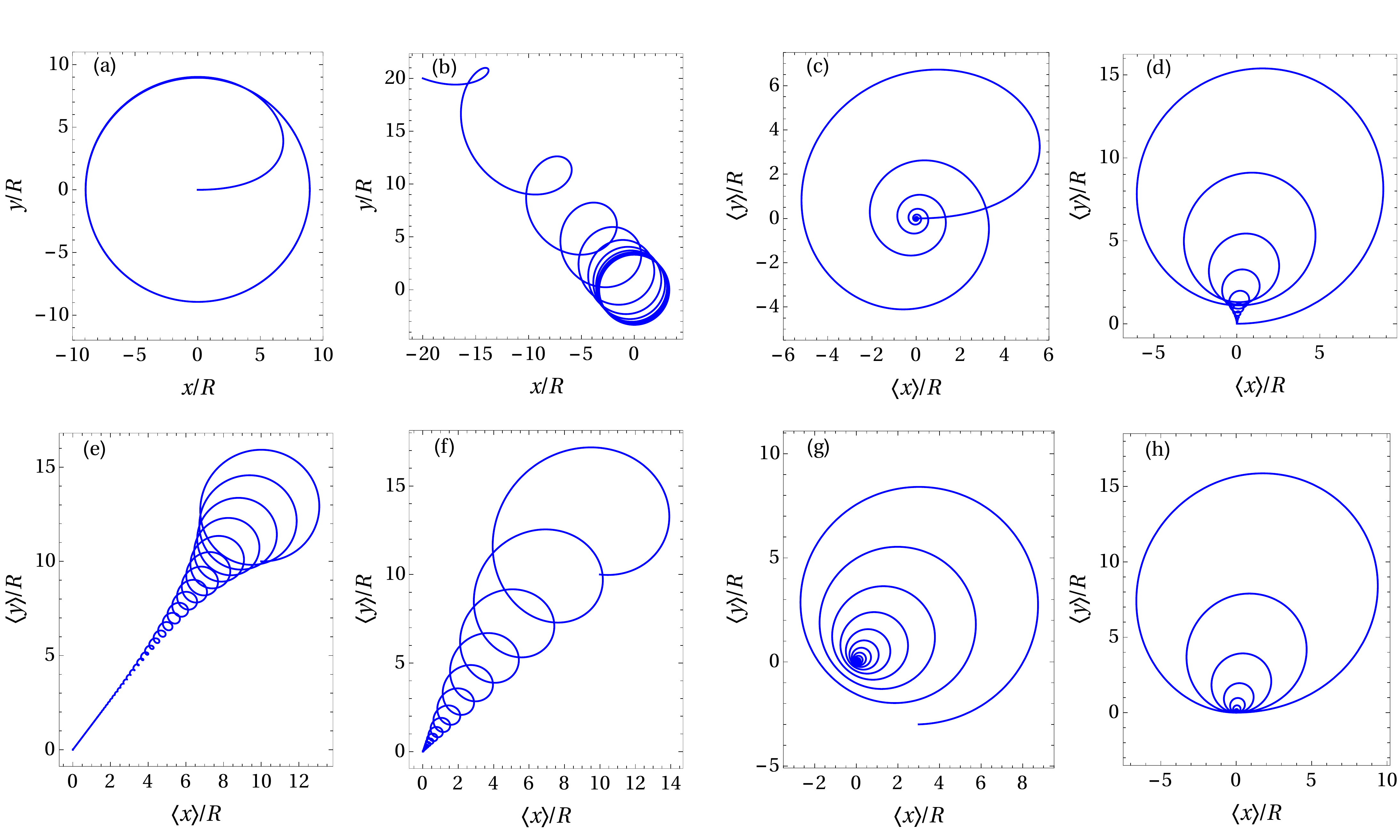}
\caption{\label{sec2_path}(Color online) (a),(b) Noise-free and (c)--(h) noise-averaged trajectories of a circle swimmer with constant self-propulsion in a constant spatial trap. In all plots, the self-propulsion force is $F_{0}=100$ and the initial angle is $\varphi_{0}=0$. The other parameters of the noise-free trajectories are in (a) $\lambda_{0}=10$, $\omega=5$, and $c_{x}=c_{y}=0$ and in (b) $\lambda_{0}=3$,  $\omega=30$ , $c_{x}=-20$, and $c_{y}=20$. For the noise-averaged trajectories, three different patterns regarding the asymptotic behavior are found based on the value of $\lambda_{0}$. The case of $\lambda_{0}>1$ is shown in (c) for $\lambda_{0}=10$, $\omega=7$, and $c_{x}=c_{y}=0$. Example mean trajectories for $\lambda_{0}<1$ are presented in (d) with $\lambda_{0}=0.7$, $\omega=10$, and $c_{x}=c_{y}=0$ and in (e) with $\lambda_{0}=0.3$, $\omega=30$, and $c_{x}=c_{y}=10$. Finally, the case of $\lambda_{0}=1$ is visualized in (f) with $\omega=20$ and $c_{x}=c_{y}=10$, in (g) with $\omega=15$, $c_{x}=3$, and $c_{y}=-3$, and in (h) with $\omega=9$ and $c_{x}=c_{y}=0$.}
\end{figure*}

We start by considering the situation of vanishing thermal noise, i.e.\ $\xi_{x}(t)=\xi_{y}(t)=\xi_{\varphi}(t)=0$. In this case, the equations of motion \eqref{Langevinx}--\eqref{Langevinphi} can be solved analytically and yield
\begin{alignat}{1}
& \frac{x\left(t\right)}{R}=e^{-\lambda_{0} D_{r}t}\left[c_{x}+f\left(D_{r}t,\omega,\varphi_{0},\lambda_{0}\right)\right]\label{II1}
 \end{alignat}
and
\begin{alignat}{1}
&\frac{y\left(t\right)}{R}=e^{-\lambda_{0} D_{r}t}\left[c_{y}+f\left(D_{r}t,\omega,\varphi_{0}-\frac{\pi}{2},\lambda_{0}\right)\right],\label{II2}
\end{alignat}
where $c_{x}=x_{0}/R$ and $c_{y}=y_{0}/R$ refer to the initial center-of-mass position $\mathbf{r}_0=(x_0,y_0)$ of the particle.  The function $f$ is specified by
\begin{alignat}{1}
&f\left(t,\omega,\varphi,\lambda\right)=\frac{F_{0}}{\lambda^{2}+\omega^{2}}\left[\lambda e^{\lambda t}\cos\left(\omega t+\varphi\right)\right. \nonumber \\
&\quad+\omega e^{\lambda t}\sin\left(\omega t+\varphi\right)\left.-\lambda\cos\left(\varphi\right)-\omega\sin\left(\varphi\right)\right]\label{II3}.
\end{alignat} 
Remarkably, after an initial regime, the microswimmer approaches a circular swimming path which is independent of the initial conditions [see Figs.\ \ref{sec2_path}(a) and \ref{sec2_path}(b)]. The analytical expression describing this limit cycle is obtained from Eqs.\ \eqref{II1} and \eqref{II2} by neglecting all exponentially decaying terms and reads
\begin{alignat}{1}
\frac{1}{R}\begin{pmatrix}
x\left(t\right)\\y\left(t\right)
\end{pmatrix}=&\frac{F_{0}}{\lambda^{2}_{0}+\omega^{2}}\Bigg[\lambda_{0}\begin{pmatrix}
\cos\left(\omega D_{r}t+\varphi_{0}\right)\\\sin\left(\omega D_{r}t+\varphi_{0}\right)
\end{pmatrix}\nonumber \\
&\quad\quad+\omega\begin{pmatrix}
\sin\left(\omega D_{r}t+\varphi_{0}\right)\\-\cos\left(\omega D_{r}t+\varphi_{0}\right)
\end{pmatrix}\Bigg].
\end{alignat} 
The radius $r_c$ of this limit cycle is
\begin{alignat}{1}
&\frac{r_{c}}{R}=\frac{F_{0}}{\sqrt{\lambda^{2}_{0}+\omega^{2}}}\label{radius}
\end{alignat}
and the particle moves along it with frequency $\omega$. The existence of the initial regime can be attributed to the effect of the spatial trap. While for relatively small ratios $\omega/\lambda_{0}$ the particle steadily approaches its final circular trajectory [see Fig.\ \ref{sec2_path}(a)], for higher ratios of circling frequency and trap strength, it takes some revolutions of the circle swimmer until the regular periodic motion is reached [see Fig.\ \ref{sec2_path}(b)]. The maximum distance the swimmer can escape from the trap center occurs at $\omega=0$ and is given by $d_\mathrm{max}=F_0/\lambda_0$. In that case, the particle orientation is fixed, and the particle moves away from the trap center on a straight line until its self-propulsion force is fully compensated by the external potential. On the other hand, for $\omega \neq 0$, the particle orientation changes continuously so that the maximum distance, which corresponds to a static situation of total force balance, cannot be reached. 

\subsection{Effect of Brownian noise}
  
In the presence of thermal fluctuations, by averaging over the noise terms the mean positions along $x$ and $y$ direction at prescribed initial conditions $\mathbf{r}_0$, $\varphi_0$ are obtained as 
  \begin{alignat}{1}
  & \frac{\left\langle x(t)\right\rangle}{R}=e^{-\lambda_{0} D_{r}t}\left[c_{x}+f\left(D_{r}t,\omega,\varphi_{0},\lambda_{0}-1\right)\right] \label{eq:mpx}
   \end{alignat}
  and
  \begin{alignat}{1}
  &\frac{\left\langle y(t)\right\rangle}{R}=e^{-\lambda_{0} D_{r}t}\left[c_{y}+f\left(D_{r}t,\omega,\varphi_{0}-\frac{\pi}{2},\lambda_{0}-1\right)\right], \label{eq:mpy}
  \end{alignat}
respectively. As shown in Figs.\ \ref{sec2_path}(c)--\ref{sec2_path}(h), the corresponding noise-averaged trajectories are spiraling curves that collapse into the trap center. The reason for this collapsing behavior is the factor $\exp(-D_{r}t)$ in Eqs.\ \eqref{eq:mpx} and \eqref{eq:mpy}, which stems from the influence of the rotational Brownian motion. While the deterministic torque $M$ causes a continuous rotation around the trap center, the random term $\xi_{\varphi}(t)$ is responsible for the reduction of the radius of successive revolutions of the noise-averaged curve. The overall effect of both torques, therefore, leads to spiraling mean trajectories collapsing into the trap center \footnote{We want to emphasize that this collapsing behavior is only observed for the mean (noise-averaged) trajectory. The swimming path of an individual particle undergoing Brownian motion will not end up perfectly in the center of the trap.}. 

Depending on the trap strength $\lambda_0$, three different cases can be distinguished. For $\lambda_{0}>1$, after relaxation of the initial conditions, a \textit{spira mirabilis} is obtained [see Fig.\ \ref{sec2_path}(c)], similar to the case without external potential \cite{vanTeeffelen2008,Kummel2013}. The corresponding asymptotic mean swimming path of the circle swimmer is given by
 \begin{alignat}{1}
  \frac{1}{R}\begin{pmatrix}\left\langle x(t)\right\rangle\\\left\langle y(t)\right\rangle\end{pmatrix}&=\frac{e^{- D_{r}t}F_{0}}{\sqrt{(\lambda_{0}-1)^2+\omega^{2}}}\begin{pmatrix}\cos\left(\omega D_{r}t+\alpha\right)\\\sin\left(\omega D_{r}t+\alpha\right)\end{pmatrix}
  \end{alignat}
with $\alpha=\varphi_{0}-\arctan(\frac{\omega}{\lambda_{0}-1})$. 
	
If the trap strength $\lambda_{0}$ is less than 1, the mean trajectory evolves into a straight line approaching the trap center as shown in Figs.\ \ref{sec2_path}(d) and \ref{sec2_path}(e). The corresponding asymptotic equations are
 \begin{alignat}{1}
  \frac{1}{R}\begin{pmatrix}\left\langle x(t)\right\rangle\\\left\langle y(t)\right\rangle\end{pmatrix}=&\frac{e^{-\lambda_{0} D_{r}t}F_{0}}{(\lambda_{0}-1)^{2}+\omega^{2}}\bigg[(1-\lambda_{0})\begin{pmatrix}\cos\varphi_{0}\\\sin\varphi_{0}\end{pmatrix}\nonumber\\&\quad+\omega\begin{pmatrix}-\sin\varphi_{0}\\\cos\varphi_{0}\end{pmatrix}\bigg]+e^{-\lambda_{0} D_{r}t}\begin{pmatrix}
 c_{x}\\c_{y}
  \end{pmatrix}.
  \end{alignat} 
  The slope of the straight line is dictated not only by the circling frequency and the trap strength but also by the initial conditions.
	
Finally, in the boundary case $\lambda_{0}=1$, the mean trajectory asymptotically becomes
    \begin{alignat}{1}
    \frac{1}{R}\begin{pmatrix}\left\langle x(t)\right\rangle\\\left\langle y(t)\right\rangle\end{pmatrix}=&e^{- D_{r}t}\frac{F_{0}}{\omega}\bigg[\begin{pmatrix}
    \sin\left(\omega D_{r}t+\varphi_{0}\right)\\-\cos\left(\omega D_{r}t+\varphi_{0}\right)\end{pmatrix}\nonumber\\&\quad\quad+\begin{pmatrix}
        -\sin\varphi_{0}\\\cos\varphi_{0}\end{pmatrix}\bigg]+e^{- D_{r}t}\begin{pmatrix}
   c_{x}\\c_{y}
    \end{pmatrix}.
    \end{alignat}
This result is visualized in Figs.\ \ref{sec2_path}(f)--\ref{sec2_path}(h) and can be referred to as ``stretched'' \textit{spira mirabilis}. The curves are obtained as a superposition of a normal \textit{spira mirabilis} and a straight line toward the trap center. If the particle starts its motion from the trap center, the corresponding mean trajectory crosses this point of minimum potential in each revolution [see Fig.\ \ref{sec2_path}(h)].

\section{Time-dependent self-propulsion}
\label{propulsion}

Next, the dynamics of a trapped circle swimmer with time-dependent self-propulsion is investigated. The self-propulsion force is modeled by
\begin{equation}
\mathbf{F}\left(t\right)=F_{0}\left[1+\cos\left(\nu t+\theta\right)\right]\hat{\mathbf{u}} \label{eq:force}
\end{equation}
with the so-called \textit{propulsion frequency} $\nu$ and the initial phase $\theta$. This approach is similar to a theoretical model that has previously been applied to untrapped microswimmers \cite{Babel2014}. In our study, the self-propulsion is implemented such that the prefactor of $\hat{\mathbf{u}}$ in Eq.\ \eqref{eq:force} remains always positive. This guarantees that the direction of propulsion is fixed in the particle's frame of reference and that the chirality of the motion does not change in time. Consequently, the Langevin equation for the translational motion is---in a nondimensionless form---given by
\begin{alignat}{1}
\frac{d}{dt}\begin{pmatrix}x\left(t\right)\\y\left(t\right)
\end{pmatrix}=\beta D&\left[F_{0}\left(1+\cos\left(\nu t+\theta\right)\right)\begin{pmatrix}\cos\left(\varphi \left(t\right)\right)\\\sin\left(\varphi \left(t\right)\right)\end{pmatrix} \right.\nonumber \\
& \left.-\lambda_{0}\begin{pmatrix}x\left(t\right)\\y\left(t\right)\end{pmatrix}\right] + \sqrt{2D}\,\begin{pmatrix}\xi_{x}\left(t\right)\\\xi_{y}\left(t\right)\end{pmatrix}, \label{eq:prop1}
\end{alignat}
while the rotational motion is governed by Eq.\ \eqref{Langevinphi} as before. Similar to the circling frequency $\omega$, we also define a dimensionless propulsion frequency $\nu{'}= \nu/D_{r}$ but omit the prime in the following.

\subsection{Results for vanishing noise}
\label{propulsionA}

Without Brownian noise, the trajectory of a circle swimmer described by the Langevin equations \eqref{eq:prop1} and \eqref{Langevinphi} is obtained as 
\begin{alignat}{1}
 \frac{x\left(t\right)}{R}=& e^{-\lambda_{0} D_{r}t}\Big[ c_{x}+f\left(D_{r}t,\omega,\varphi_{0},\lambda_{0}\right) \nonumber \\
 & +\frac{1}{2}f\left(D_{r}t,\omega+\nu,\varphi_{0}+\theta,\lambda_{0}\right) \nonumber \\
  & +\frac{1}{2}f\left(D_{r}t,\omega-\nu,\varphi_{0}-\theta,\lambda_{0}\right)\Big],  \label{eq:zero1}\\
 \frac{y\left(t\right)}{R} =& e^{-\lambda_{0} D_{r}t}\Big[c_{y}+f\left(D_{r}t,\omega,\varphi_{0}-\frac{\pi}{2},\lambda_{0}\right) \nonumber \\
 & +\frac{1}{2}f\left(D_{r}t,\omega+\nu,\varphi_{0}+\theta-\frac{\pi}{2},\lambda_{0}\right) \nonumber \\
  & +\frac{1}{2}f\left(D_{r}t,\omega-\nu,\varphi_{0}-\theta-\frac{\pi}{2},\lambda_{0}\right)\Big] \label{eq:zero2}
 \end{alignat}
with the function $f(t,\omega,\varphi,\lambda)$ as defined in Eq.\ \eqref{II3}. An example trajectory based on Eqs.\ \eqref{eq:zero1} and \eqref{eq:zero2} is shown in Fig.\ \ref{Flo:1}(a). There and in all following figures, for simplicity we set $c_{x}=c_{y}=0$, i.e.\ the initial position of the particle is always in the center of the trap. When all exponentially decaying terms in Eqs.\ \eqref{eq:zero1} and \eqref{eq:zero2} are neglected, the expression reduces to
\begin{alignat}{1}
&\frac{1}{R}\begin{pmatrix}
x\left(t\right)\\y\left(t\right)
\end{pmatrix}=\frac{F_{0}}{\lambda^{2}_{0}+\omega^{2}} \nonumber \\
&\times\Bigg[\lambda_{0}\begin{pmatrix}
\cos\left(\omega D_{r}t+\varphi_{0}\right)\\\sin\left(\omega D_{r}t+\varphi_{0}\right)
\end{pmatrix}+\omega\begin{pmatrix}
\sin\left(\omega D_{r}t+\varphi_{0}\right)\\-\cos\left(\omega D_{r}t+\varphi_{0}\right)
\end{pmatrix}\nonumber \\
&+\frac{\lambda_{0}}{2}\begin{pmatrix}
\cos\Big((\omega+\nu) D_{r}t+\varphi_{0}+\theta\Big)\\\sin\Big((\omega+\nu) D_{r}t+\varphi_{0}+\theta\Big)
\end{pmatrix}\nonumber \\
&+\frac{(\omega+\nu)}{2}\begin{pmatrix}
\sin\Big((\omega+\nu) D_{r}t+\varphi_{0}+\theta\Big)\\-\cos\Big((\omega+\nu) D_{r}t+\varphi_{0}+\theta\Big)
\end{pmatrix}\nonumber \\
&+\frac{\lambda_{0}}{2}\begin{pmatrix}
\cos\Big((\omega-\nu) D_{r}t+\varphi_{0}-\theta\Big)\\\sin\Big((\omega-\nu) D_{r}t+\varphi_{0}-\theta\Big)
\end{pmatrix}\nonumber \\
&+\frac{(\omega-\nu)}{2}\begin{pmatrix}
\sin\Big((\omega-\nu) D_{r}t+\varphi_{0}-\theta\Big)\\-\cos\Big((\omega-\nu) D_{r}t+\varphi_{0}-\theta\Big)
\end{pmatrix}\Bigg].\label{eq:zerovec}
\end{alignat} 
Corresponding trajectories without initial regimes are presented in Figs.\ \ref{Flo:1}(b)--\ref{Flo:1}(d).

If the ratio of the frequencies $\nu$ and $\omega$ is rational, the swimming paths according to Eq.\ \eqref{eq:zerovec} are closed rosette-like curves. For $\nu \neq \omega$, their period is determined by $T_p=\frac{2\pi}{D_{r}}\,\mathrm{LCM}\left(\frac{1}{\omega},\frac{1}{\omega+\nu},\frac{1}{\left|\omega-\nu\right|}\right)$, where $\mathrm{LCM}$ denotes the lowest common multiple. In the special case $\nu=\omega$, the period is $T_p=\frac{2\pi}{D_{r}\omega}$. If the ratio $\nu/\omega$ is irrational, the trajectory will never close and eventually fill a circular area around the trap center. Therefore, we only consider situations where $\nu/\omega$ is rational in the following.  

\begin{figure}[tb]
\centering
\includegraphics[width = \columnwidth]{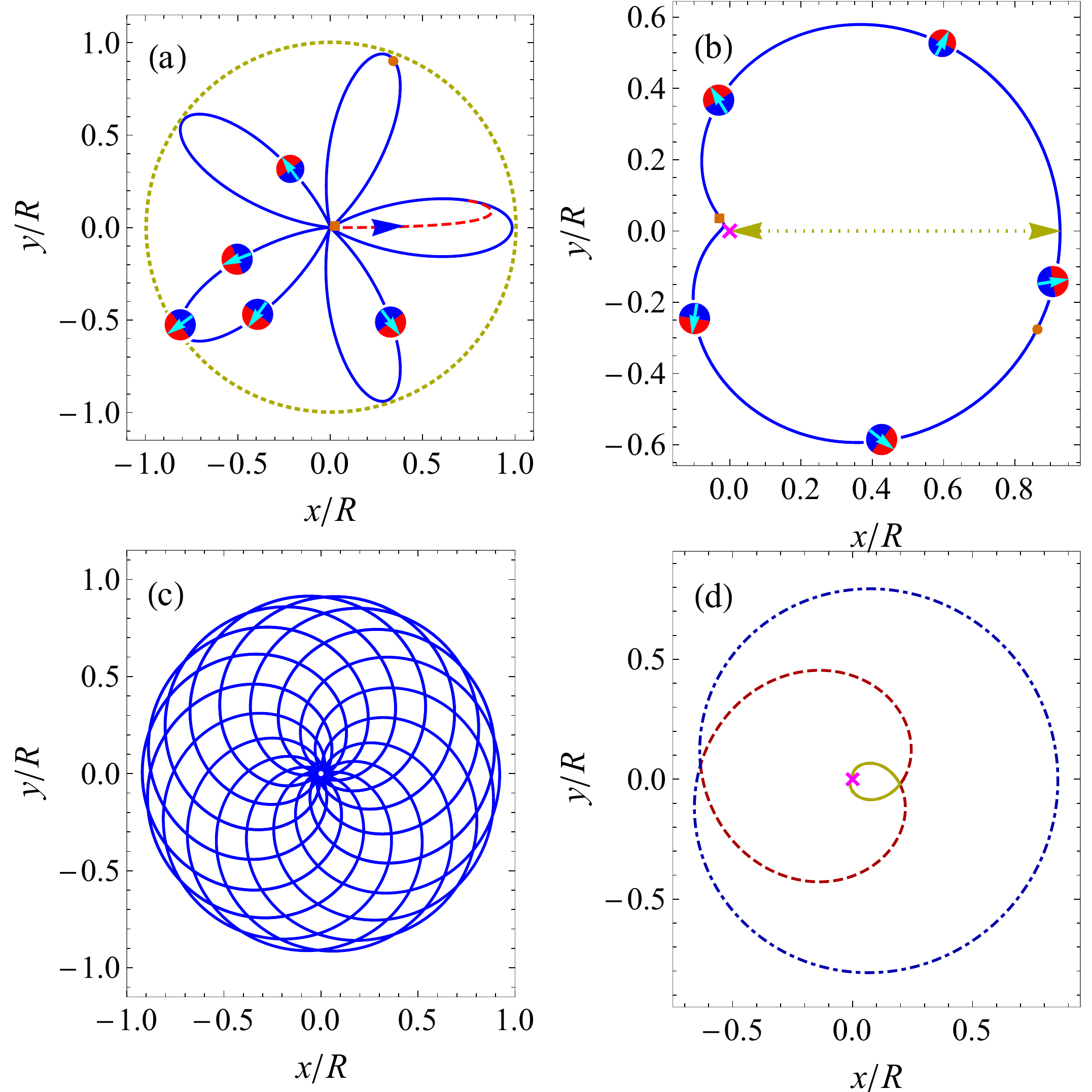}
\caption{\label{Flo:1}(Color online) Noise-free trajectories of a circle swimmer with temporally varying self-propulsion force in a constant spatial trap. While the parameters $F_{0}=1$, $\lambda_{0}=2$, and $\theta=\varphi_{0}=0$ are the same in all plots, the values of the circling frequency $\omega$ and the propulsion frequency $\nu$ are varied as follows: (a)  $\omega=0.1$ and
$\nu=0.5$, (b) $\omega=\nu=0.7$, (c)  $\omega=0.75$ and
$\nu=0.7$, and (d) $\omega=1.2$ and $\nu=0.4$. The behavior for short times, which depends on the initial conditions, is only shown in (a), where it is visualized by the dashed (red) curve. Furthermore, in (a) and (b) the particle orientation at various points of the trajectory is illustrated by the sketched Janus particles whose direction of propulsion is represented by arrows. The orange bullet symbol marks the point where the self-propulsion is maximal, and the square symbol refers to the position with minimal self-propulsion.
 The radius of the dotted circle in (a) and the dotted line in (b) indicate the maximum distance from the trap center which the particle can reach. This distance is analyzed in detail in Fig.~\ref{6n}. 
 The swimming path in (d) is a single-petal curve with two inner petals. Different line styles (and colors) are used to illustrate the individual inner petals. The magenta cross in (b) and (d) indicates the trap center.}
\end{figure}

The evidence of the presence of the trap is manifested in the initial motion of the particle before it reaches its periodic trajectory. In particular, the spatial trap dictates the position of the swimming path. For instance, in the trajectory of Fig.\ \ref{Flo:1}(a), the
particle starts its motion from the origin, but its initial motion---the dashed (red-colored) path---deviates from its following periodic rosette-like path. Each ``petal'' is formed by the interplay of the time-dependent self-propulsion force, the external potential, and the torque. Precisely, in Fig.\ \ref{Flo:1}(a) the circle swimmer starts in the center of the spatial trap with the maximum value of the self-propulsion force due to the initial conditions. Subsequently, it escapes from the trap center while the self-propulsion decreases. Simultaneously, the swimmer rotates as determined by the circling frequency. Meanwhile, the trap force acting on the particle increases since it is proportional to the distance from the center. At the outermost point of the initial petal [dashed red curve in Fig.\ \ref{Flo:1}(a)], the trap force becomes so strong that it is able to pull the particle back toward the center. Returning to the trap center
continues roughly until the propulsion force adopts its minimum value. Shortly after that, with increasing self-propulsion, the particle once more starts to run away from the center before the latter is actually reached. Since, meanwhile, the torque is also rotating the circle swimmer, the next petal begins to form. For all petals that are not influenced by the initial regime anymore, the maximum distance from the trap center is reached slightly after the self-propulsion force adopts its maximum value. (This point in time is indicated by the orange bullet in Fig.\ \ref{Flo:1}(a).) Apart from that, the process corresponds to that described in detail for the initial petal.  

Although most of the occurring trajectories have an $m$-fold rotational symmetry around the trap center with $m\geq 2$---similar to the example trajectories in Figs.~\ref{Flo:1}(a) and \ref{Flo:1}(c)---, under certain conditions a circle swimmer with temporally varying self-propulsion in a symmetric two-dimensional spatial trap can also undergo swimming paths that are not symmetric around the center of the trap, even after relaxation of the initial conditions.
For instance, for $\omega=\nu$ the particle always
moves on a single-petal heart-shaped path as shown in Fig.\
\ref{Flo:1}(b). The reason is the synchronous oscillation of the self-propulsion force and the particle orientation, which leads to the formation of the second petal exactly on top of the first one. 
Generally, trajectories that are asymmetric with respect to the trap center occur only for $\omega=n\nu$ with $n=1, 2, 3, \ldots$ since in these cases, during one period of the self-propulsion force, the circle swimmer completes $n$ full revolutions. This leads to a single-petal trajectory with $n-1$ inner petals [cf.\ Fig.\ \ref{Flo:1}(d)]. It should be mentioned that the inner petals are qualitatively different from the petals of a symmetric rosette-like trajectory as in Fig.\ \ref{Flo:1}(a) since they are in general not created one after the other by the swimmer. 

\begin{figure}[tb]
\centering
\includegraphics[width=\columnwidth]{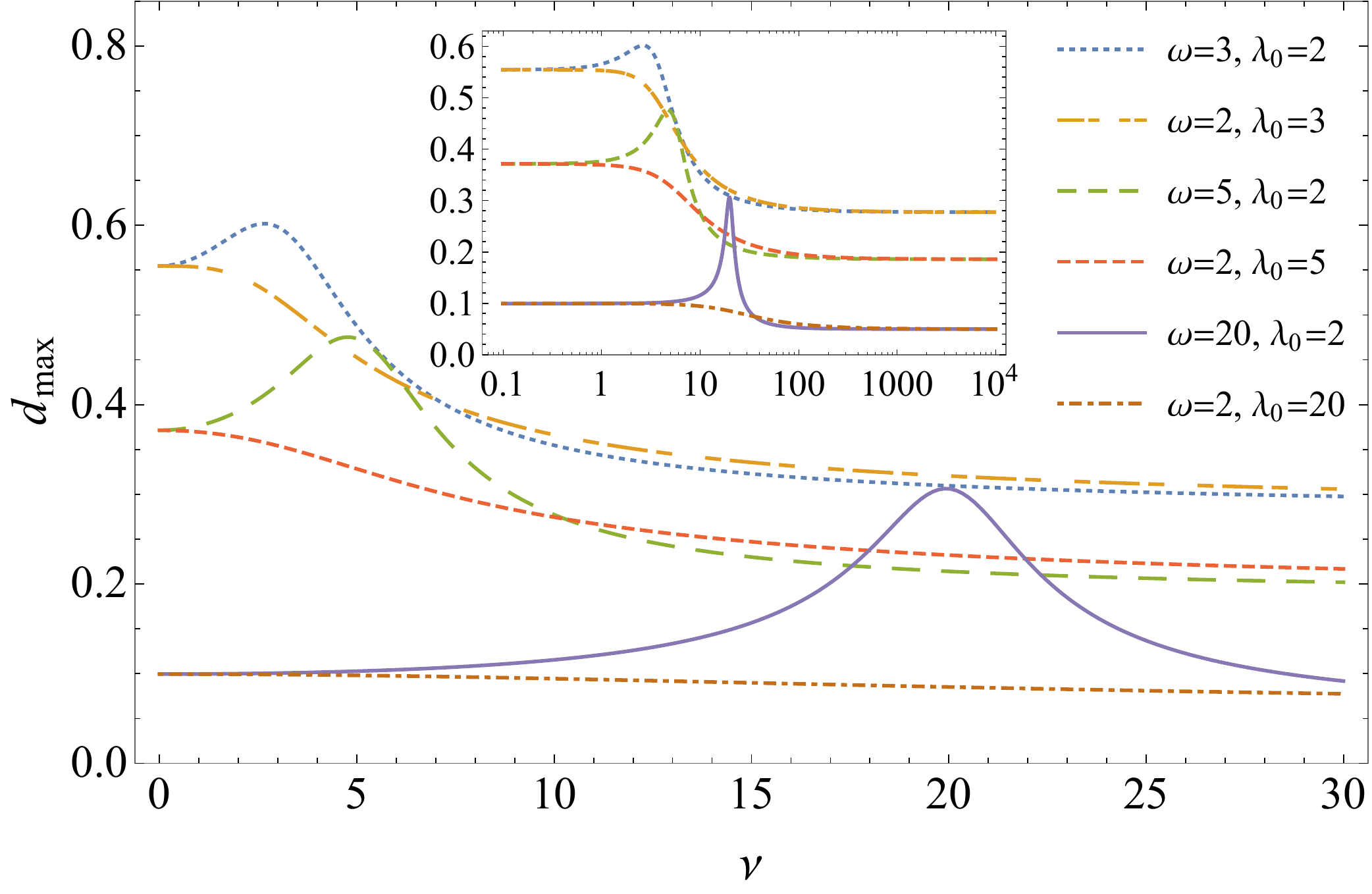}
\caption{\label{6n}(Color online) Maximum distance $d_\mathrm{max}$ from the trap center of a noise-free circle swimmer with temporally varying self-propulsion in a constant harmonic trap as a function
of $\nu$ for
$F_{0}=1$ and $\theta=0$. Any effect from the initial behavior of the particle has been neglected. The maximum value of $d_\mathrm{max}$ occurs at $\nu=2.7$ for the dotted blue curve, at $\nu=4.8$ for the dashed green curve, and at $\nu=19.9$ for the solid purple curve. The maximum of the other three curves is at $\nu=0$. In the inset, the asymptotic behavior for large values of $\nu$ is visualized.}
\end{figure}

An interesting quantity to study is the maximum distance $d_\mathrm{max}=\mathrm{max}\left(\sqrt{x^2+y^2}/R\right)$ from the center of the trap which a particle
can reach on its periodic trajectory. For instance, the maximum distance in Fig.\ \ref{Flo:1}(a) is the radius of the dotted circle, and in Fig.\ \ref{Flo:1}(b) $d_\mathrm{max}$ is indicated by the dotted line. A detailed analysis of this maximum distance from the trap center is provided in Fig.\ \ref{6n}. Interestingly, a pronounced resonance situation is observed, which is determined by the propulsion
frequency $\nu$, the circling frequency $\omega$, and also the trap strength $\lambda_0$. By varying
$\omega$ and $\lambda_{0}$ and plotting $d_\mathrm{max}$ as a function of $\nu$, it is concluded that the maximum escape distance
of the particle roughly occurs for $\nu^{2}=\omega^{2}-\lambda_{0}^{2}/2$ if the right-hand side of this equation is positive (see Fig.\ \ref{6n}). Thus, to achieve an optimal swimming strategy in order to explore a maximum spatial range around the trap center, the particle has to adapt its propulsion frequency to a given circling frequency and trap strength accordingly. For $\omega^{2}-\lambda_{0}^{2}/2<0$, the maximum value of $d_\mathrm{max}$ is always at $\nu=0$. The strength of the self-propulsion force has no effect on the resonance situation, apart from changing the value of  $d_\mathrm{max}$.

By neglecting the initial regime and setting $\theta=0$, the asymptotic behavior of $d_\mathrm{max}$ in the limits of $\nu\rightarrow 0$ and $\nu\rightarrow\infty$ is obtained as
\begin{equation}
\underset{\nu\rightarrow 0}{\lim} \, d_\mathrm{max} =\frac{2F_{0}}{\sqrt{\lambda_{0}^{2}+\omega^{2}}}
\end{equation}
and
\begin{equation}
\underset{\nu\rightarrow\infty}{\lim} d_\mathrm{max} =\frac{F_{0}}{\sqrt{\lambda_{0}^{2}+\omega^{2}}},
\end{equation}
respectively. Obviously, the limit for large values of $\nu$ is exactly half of the limit $\nu\rightarrow 0$. If the self-propulsion force $F_{0}$ is kept at a constant value, exchanging the values of the trap strength $\lambda_{0}$ and the circling frequency $\omega$ does not alter the limit values although the swimming paths are different. This is illustrated by the curves in the inset of Fig.\ \ref{6n}.

\subsection{Effect of Brownian noise}
\label{propulsionB}

\begin{figure}[tb]
\centering
\includegraphics[width = \columnwidth]{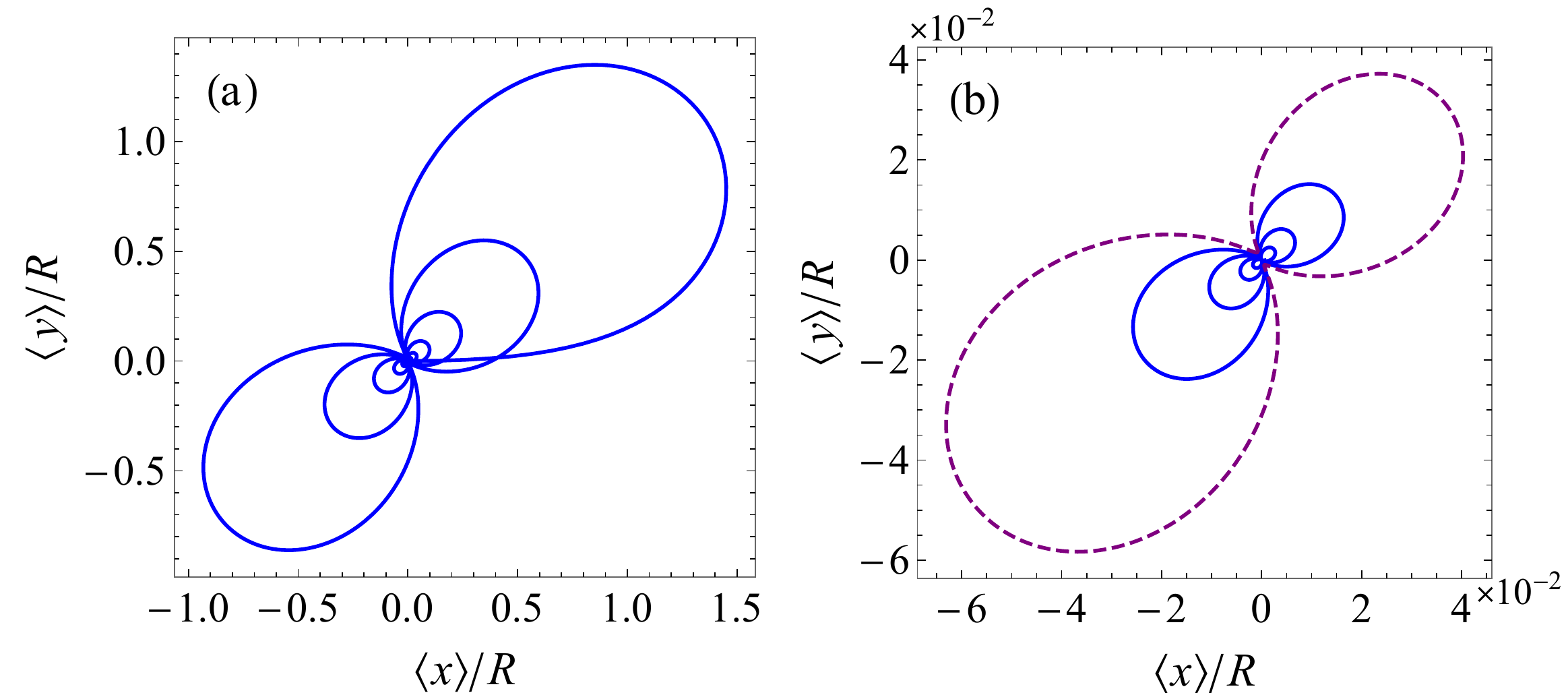}
	\caption{\label{Flo:9}(Color online) Noise-averaged trajectory of a circle swimmer with temporally oscillating self-propulsion force in a constant
	harmonic trap. The parameters are $F_{0}=50$,	$\lambda_{0}=50$, $\omega=14$, 
	$\nu=7$, $\theta=3\pi/2$, and $\varphi_{0}=0$. In (a) the complete trajectory including the initial regime is shown while the close-up view in (b) clearly illustrates the self-similarity of the curve. The mean swimming path during one period is indicated by the dashed purple line.}
\end{figure}

If the Brownian noise terms in Eqs.\ \eqref{eq:prop1} and \eqref{Langevinphi} are taken into account, the mean positions of the circle swimmer along $x$ and $y$ direction are obtained as
\begin{alignat}{1}
\left\langle \frac{x\left(t\right)}{R}\right\rangle & =e^{-\lambda_{0} D_{r}t}\Big[c_{x}+f\left(D_{r}t,\omega,\varphi_{0},\lambda_{0}-1\right)\nonumber \\
 & +\frac{1}{2}f\left(D_{r}t,\omega+\nu,\varphi_{0}+\theta,\lambda_{0}-1\right) \nonumber \\
  & +\frac{1}{2}f\left(D_{r}t,\omega-\nu,\varphi_{0}-\theta,\lambda_{0}-1\right)\Big] \label{IIB1}
 \end{alignat}
and
\begin{alignat}{1}
&\left\langle \frac{y\left(t\right)}{R}\right\rangle =e^{-\lambda_{0} D_{r}t}\Big[c_{y}+f\left(D_{r}t,\omega,\varphi_{0}-\frac{\pi}{2},\lambda_{0}-1\right) \nonumber \\
 &\quad\quad\quad +\frac{1}{2}f\left(D_{r}t,\omega+\nu,\varphi_{0}+\theta-\frac{\pi}{2},\lambda_{0}-1\right) \nonumber \\
  &\quad\quad\quad +\frac{1}{2}f\left(D_{r}t,\omega-\nu,\varphi_{0}-\theta-\frac{\pi}{2},\lambda_{0}-1\right)\Big], \label{IIB2}
 \end{alignat}
respectively, based on the function $f(t,\omega,\varphi,\lambda)$ defined in Eq.\ \eqref{II3}. After times long enough for the particle to abandon its initial
regime, the mean trajectory does not change in time except for a factor $\exp(-D_{r}t)$, which originates from the random torque. 
Thus, self-similar curves collapsing toward the trap center are obtained if the initial behavior is excluded. A shrunk version of the mean trajectory is repeated after the period $T_p=\frac{2\pi}{D_{r}}\,\mathrm{LCM}\left(\frac{1}{\omega},\frac{1}{\omega+\nu},\frac{1}{\left|\omega-\nu\right|}\right)$. An example of such a noise-averaged trajectory is shown in Fig.\ \ref{Flo:9}. The number of petals is the same as in the corresponding noise-free trajectory and also their shape is similar. However, due to the rotational Brownian noise, subsequent petals become smaller and lie inside the petals formed during the preceding periods. 

In the following, we focus on the mean square displacement, which is the standard quantity to characterize Brownian dynamics. As we restrict our investigation to situations where the particle starts in the origin, which coincides with the center of the trap, the mean square displacement is identical to the second moment $\left(\left\langle x^{2}\left(t\right)\right\rangle+\left\langle y^{2}\left(t\right)\right\rangle\right)/R^{2}$. After the initial regime, the mean square displacement is given by 
\begin{alignat}{1}  
&\underset{t\rightarrow\infty}{\lim} \frac{\left\langle x^{2}\right\rangle +\left\langle y^{2}\right\rangle }{R^{2}} =\frac{\big(2C\lambda_{0}-B\nu\big)}{4\lambda_{0}^{2}+\nu^{2}}\cos\big(\nu D_r t+\theta\big)\nonumber \\
&\quad+\frac{\big(2B\lambda_{0}+C\nu\big)}{4\lambda_{0}^{2}+\nu^{2}}\sin\big(\nu D_r t+\theta\big)\nonumber \\
&\quad+\frac{\big(A\lambda_{0}-B\nu\big)}{4\lambda_{0}^{2}+4\nu^{2}}\cos\big(2 \nu D_r t+2\theta\big)\nonumber \\
&\quad+\frac{\big(B\lambda_{0}+A\nu\big)}{4\lambda_{0}^{2}+4\nu^{2}}\sin\big(2 \nu D_r t+2\theta\big)\nonumber \\
&\quad+\frac{F_{0}^{2}}{\lambda_{0}}\frac{\lambda_{0}+1}{\left(\lambda_{0}+1\right)^{2}+\omega^{2}}+\frac{A}{4\lambda_{0}}+\frac{8}{3\lambda_{0}}\,,
\label{eq:msdlong}
\end{alignat}
where the constants $A$, $B$, and $C$ are defined as
\begin{alignat}{1}
& A=\frac{F_{0}^{2}\big(\lambda_{0} +1\big)}{\left(\lambda_{0} +1\right)^{2}+\left(\omega-\nu\right)^{2}}+\frac{F_{0}^{2}\big(\lambda_{0} +1\big)}{\left(\lambda_{0}+1\right)^{2}+\left(\omega+\nu\right)^{2}}\,,
\end{alignat}
\begin{alignat}{1}
& B=\frac{F_{0}^{2}\big(\nu-\omega\big)}{\left(\lambda_{0} +1\right)^{2}+\left(\omega-\nu\right)^{2}}+\frac{F_{0}^{2}\big(\omega+\nu\big)}{\left(\lambda_{0}+1\right)^{2}+\left(\omega+\nu\right)^{2}}\,,
\end{alignat}
and
\begin{alignat}{1}
C &=\frac{2F_{0}^{2}\big(\lambda_{0} +1\big)}{\left(\lambda_{0}+1\right)^{2}+\omega^{2}}+A\,,
\end{alignat}
respectively. 

\begin{figure}[tb]
\centering
\includegraphics[width = \columnwidth]{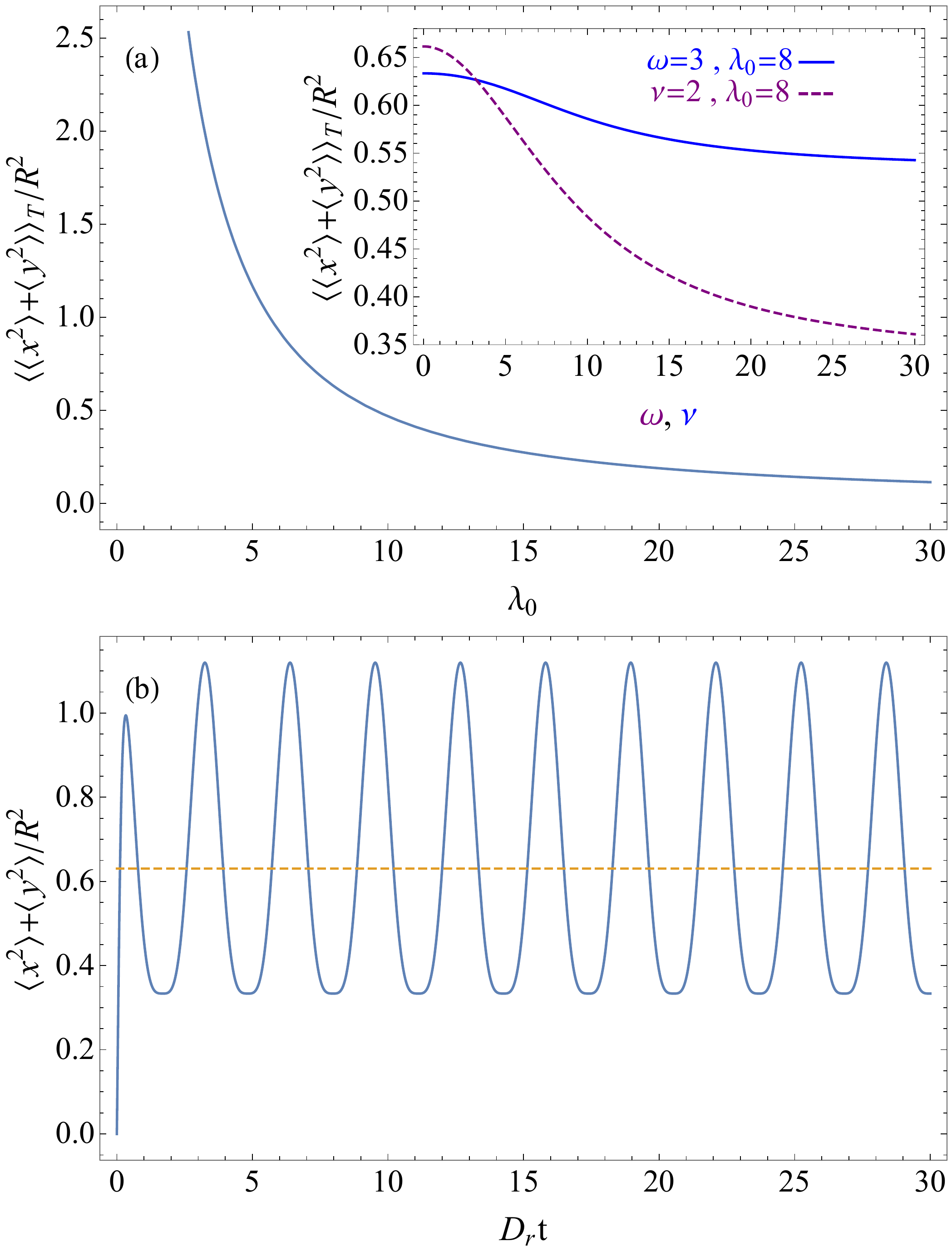}
\caption{\label{msd}(Color online) (a) Time-averaged mean square displacement of a Brownian circle swimmer with temporally varying self-propulsion in a constant harmonic trap as a function of the trap strength $\lambda_{0}$. The parameters are $F_{0}=4$, $\omega=3$, $\nu=2$, and $\theta=0$. The inset shows the same quantity as a function of the frequencies $\omega$ and $\nu$. All curves are based on the analytical result in Eq.\ \eqref{eq:11}. 
(b) Full time dependence of the mean square displacement according to Eq.\ \eqref{eq:msd} in the Appendix for $\lambda_{0}=8$ and the other parameters being the same as in (a). The dashed orange line represents the time-averaged value for the chosen parameters.}
\end{figure}

While in Eq.\ \eqref{eq:msdlong} all terms decaying exponentially in time have been neglected, an additional time average---indicated by the notation $\left\langle \ldots\right\rangle _{T}$ in the following---leads to the expression 
\begin{equation}
\left\langle \frac{\left\langle x^{2}\right\rangle +\left\langle y^{2}\right\rangle }{R^{2}}\right\rangle _{T}=\frac{F_{0}^{2}}{\lambda_{0}}\frac{\lambda_{0}+1}{\left(\lambda_{0}+1\right)^{2}+\omega^{2}}+\frac{A}{4\lambda_{0}}+\frac{8}{3\lambda_{0}}.\label{eq:11}
\end{equation} 
This result is visualized in Fig.\ \ref{msd}(a), where the dependence on the potential strength $\lambda_{0}$ is shown. In the inset, the time-averaged mean square displacement is plotted as a function of the frequencies $\omega$ and $\nu$, respectively. Increasing the values of $\lambda_{0}$, $\omega$, or $\nu$ always leads to a decrease of the time-averaged mean square displacement. 
 The explicit time dependence of the mean square displacement is shown in Fig.\ \ref{msd}(b). The solid blue curve is based on the full analytical solution, which is provided in Eq.\ \eqref{eq:msd} in the Appendix. As a function of time, the curve oscillates with the frequency $\nu$ around the mean value determined by Eq.\ \eqref{eq:11}.
 
\begin{figure}[tb]
\centering
\includegraphics[width=\columnwidth]{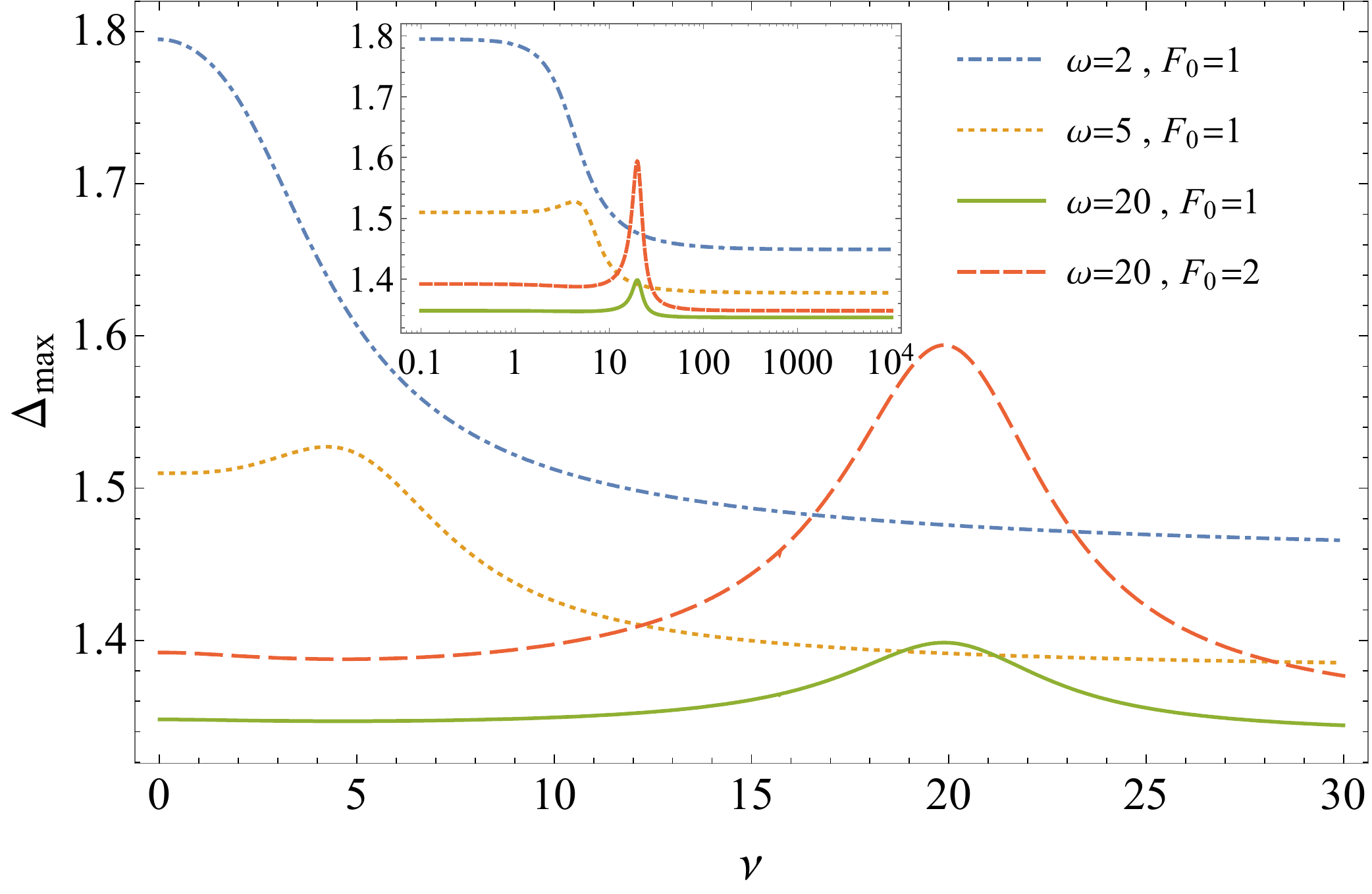}
\caption{\label{max_msp}(Color online) Maximum mean square displacement $\Delta_\mathrm{max}$ of a Brownian circle swimmer with temporally varying self-propulsion in a constant harmonic trap as a function
of the propulsion frequency $\nu$. The parameters that are not specified in the figure are $\lambda_{0}=2$ and $\varphi_{0}=\theta=0$. Any influence of the initial regime has been neglected. The maximum of $\Delta_\mathrm{max}$ occurs at $\nu=0$ for the dot-dashed blue curve, at $\nu=4.3$ for the dotted orange curve, and at $\nu=19.9$ for both the solid green and the dashed red curves. The inset visualizes the asymptotic behavior for large values of $\nu$.}
\end{figure}

On top of the general behavior of the mean square displacement, special focus is directed at the maximum value $\Delta_\mathrm{max}=\mathrm{max}\left(\left(\left\langle x^{2}\right\rangle+\left\langle y^{2}\right\rangle\right)/R^{2}\right)$. Similar to the maximum escape distance $d_\mathrm{max}$ in the case of vanishing noise, a resonance situation is found, which is defined by the system frequencies and the strength of the spatial trap. This resonance is visualized in Fig.\ \ref{max_msp}, where $\Delta_\mathrm{max}$ is plotted as a function of the propulsion frequency $\nu$. As illustrated by the solid green and the dashed red curves, the self-propulsion strength only affects the maximum value of $\Delta_\mathrm{max}$ but does not shift the resonance frequency.  
The semilogarithmic plot in the inset of Fig.\ \ref{max_msp} reveals the asymptotic behavior of $\Delta_\mathrm{max}$ for very large values of $\nu$.
The limit cases $\nu\rightarrow 0$ and $\nu\rightarrow\infty$ have also been calculated analytically from Eq.\ \eqref{eq:msdlong} and yield for $\theta=0$ 
\begin{equation}
\underset{\nu\rightarrow 0}{\lim} \, \Delta_\mathrm{max} =4\frac{F_{0}^{2}}{\lambda_{0}}\frac{\lambda_{0}+1}{\left(\lambda_{0}+1\right)^{2}+\omega^{2}}+\frac{8}{3\lambda_{0}}\label{nu0}
\end{equation}
and
\begin{equation}
\underset{\nu\rightarrow\infty}{\lim}\Delta_\mathrm{max}=\frac{F_{0}^{2}}{\lambda_{0}}\frac{\lambda_{0}+1}{\left(\lambda_{0}+1\right)^{2}+\omega^{2}}+\frac{8}{3\lambda_{0}},\label{nuInfinity}
\end{equation}
respectively.

\section{Time-dependent harmonic potential}
\label{potential}

\begin{figure}[tb]
\centering
\includegraphics[width = \columnwidth]{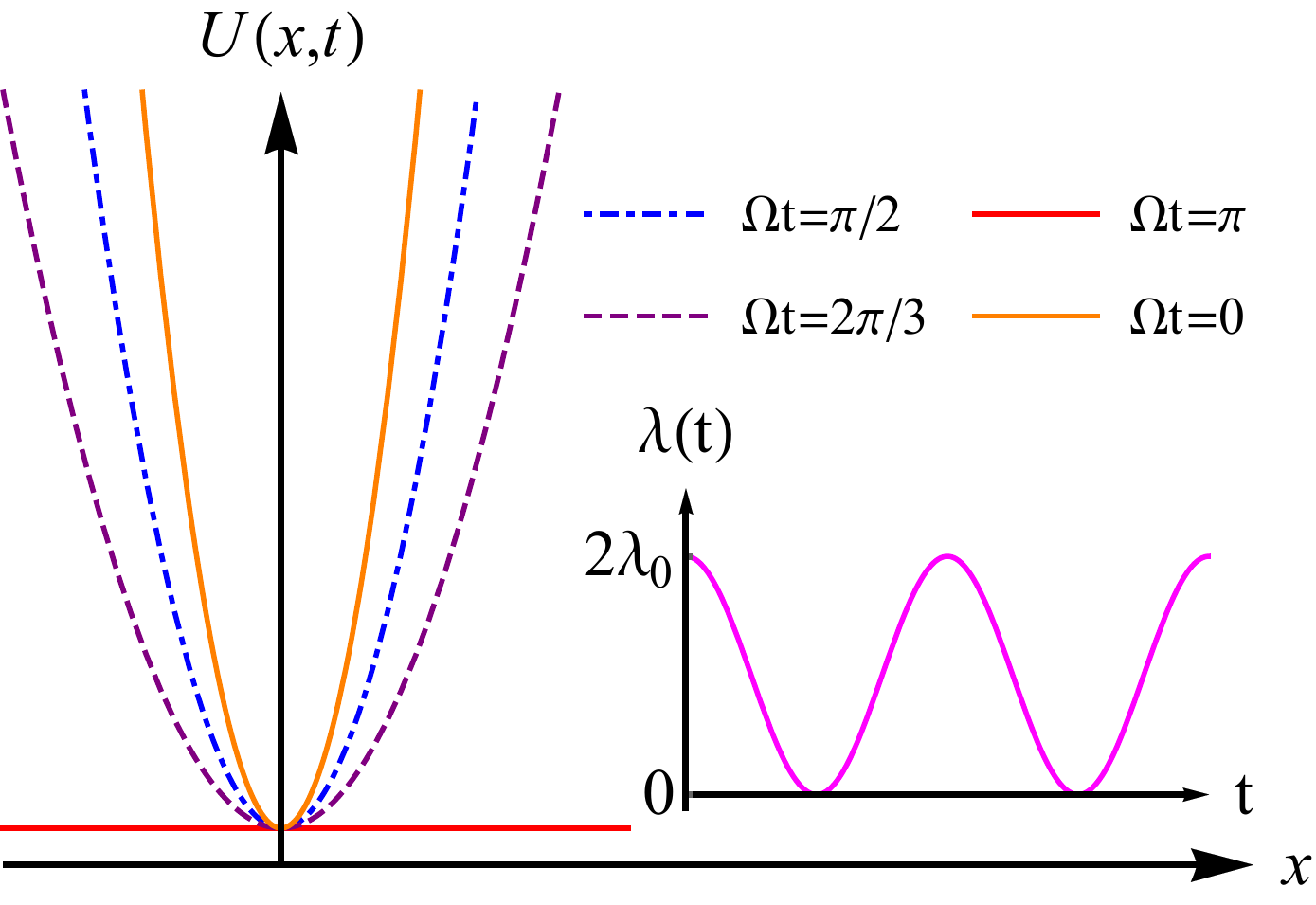}
\caption{\label{fig2}(Color online) Sketch of the time-dependent harmonic potential along the $x$ direction for different times $\Omega t$. The potential oscillates between the solid orange curve and the horizontal red line. The inset shows the behavior of the trap strength $\lambda$ versus time.}
\end{figure}

Now, we keep the self-propulsion force fixed at the value $F_{0}$ and instead consider a time dependence in the strength of the spatial trap, given by 
\begin{equation}
\lambda\left(t\right)=\lambda_{0}\left(1+\,\cos\left(\Omega t\right)\right).
\label{eq:trap}
\end{equation}
This situation constitutes a simple reference case for a time-dependent confinement. 
The constant term in Eq.~\eqref{eq:trap} guarantees that the potential is always attractive, i.e.\ confining. Thus, the particle is under the influence of a temporally opening-closing trap as sketched in Fig.\ \ref{fig2}. Accordingly, the frequency $\Omega$ is referred to as \textit{breathing frequency} \cite{Lowen2009}.

The translational Langevin equation adapted to this new situation reads in a nondimensionless form 
\begin{alignat}{1}
&\frac{d}{dt}\begin{pmatrix}x\left(t\right)\\y\left(t\right)
\end{pmatrix}=\beta D\left[F_{0}\begin{pmatrix}\cos\left(\varphi \left(t\right)\right)\\\sin\left(\varphi \left(t\right)\right)\end{pmatrix} \right.\nonumber \\
&\quad \left.-\lambda_{0}\left(1+\,\cos\left(\Omega t\right)\right)\begin{pmatrix}x\left(t\right)\\y\left(t\right)\end{pmatrix}\right] + \sqrt{2D}\,\begin{pmatrix}\xi_{x}\left(t\right)\\\xi_{y}\left(t\right)\end{pmatrix}. \label{xLangOscTrap}
\end{alignat}
Again, as already done for the circling frequency $\omega$ and the propulsion frequency $\nu$, we define a dimensionless breathing frequency $\Omega^{'}=\Omega/D_{r}$ but omit the prime in the following. Since a full analytical solution of Eq.~\eqref{xLangOscTrap} is not available, most of the results in this section were obtained numerically.

\subsection{Results for vanishing noise}

Similar to the case of a circle swimmer with time-dependent self-propulsion in a constant harmonic trap as studied in Sec.\ \ref{propulsion}, subsequent to an initial regime the noise-free trajectories obtained from the Langevin equations \eqref{xLangOscTrap} and \eqref{Langevinphi} are closed rosette-like curves for rational ratios of the frequencies $\Omega$ and $\omega$.  This time, the trajectories are governed by the interplay between the circle swimming and the temporal behavior of the spatial trap. Correspondingly, the period is defined by $T_p=\frac{2\pi}{D_{r}}\,\mathrm{LCM}\left(\frac{1}{\omega},\frac{1}{\Omega}\right)$. As before, most of the trajectories consist of several petals and have an initial regime that is determined by the initial conditions $\mathbf{r}_0$, $\varphi_0$. 

\begin{figure}[tb]
\centering
\includegraphics[width = \columnwidth]{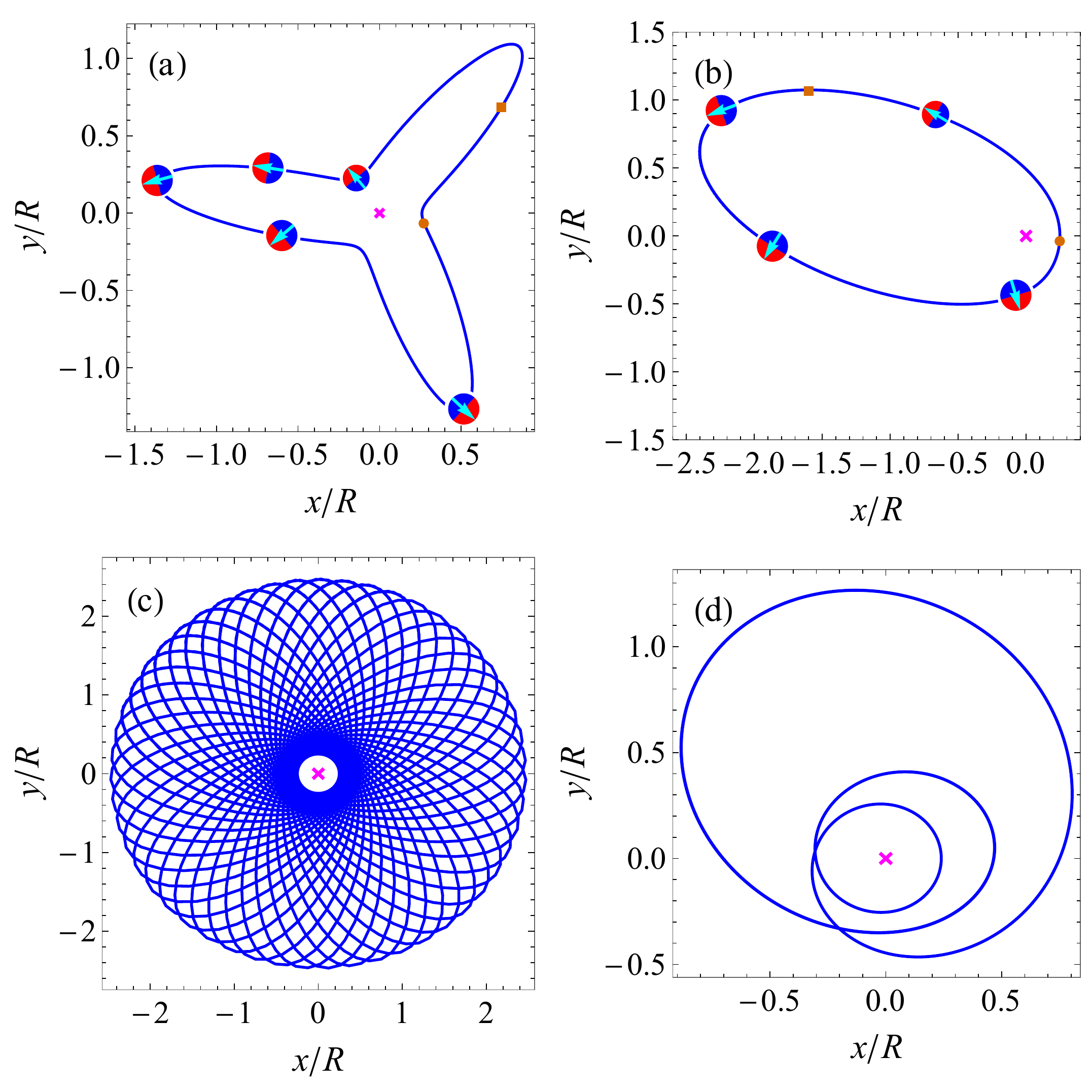}
\caption{\label{sec4}(Color online) Noise-free trajectories of a circle swimmer with constant self-propulsion in a temporally varying harmonic trap. The parameters $F_{0}=1$, $\lambda_{0}=2$, and $\varphi_{0}=0$ are the same in all plots. The values of the breathing frequency $\Omega$ and the circling frequency $\omega$ are varied as follows: (a) $\Omega=1.5$ and $\omega=0.5$, (b) $\Omega=\omega=0.5$, (c) $\Omega=0.5$ and $\omega=0.51$, and (d) $\Omega=0.4$ and $\omega=1.2$. All plots show the periodic trajectories for long times after relaxation of the initial conditions. The magenta cross represents the center of the trap. In (a) and (b), the particle orientation at various points of the trajectory is visualized by the sketched Janus particles whose direction of propulsion is indicated by arrows. The orange square symbol marks the point where the trap strength is minimal, and the bullet symbol refers to the maximum potential.}
\end{figure}

Some example trajectories are shown in Fig.\ \ref{sec4}. For reasons of clarity, the initial behavior of the circle swimmer has been removed from the plots. The formation of the swimming path is exemplarily described based on Fig.\ \ref{sec4}(a): In that case, three petals arise since the breathing frequency $\Omega = 1.5$ is three times as large as the circling frequency $\omega=0.5$. Thus, during one full rotation of the particle, the trap completes three oscillations, which each lead to the formation of one petal. The particle reaches the outermost point of a petal some time after the potential is minimal, and the turning points that are closest to the trap center correspond to times when the potential has just started to decrease again from its maximum value. The sketched Janus particles inside the figure indicate the particle orientation at the respective  points of the trajectory.  As the relation $\Omega = 3 \omega$ holds for the situation shown in Fig.\ \ref{sec4}(a), the fourth petal is formed exactly on top of the first one so that the periodic trajectory arises.  

\begin{figure}[tb]
\centering
\includegraphics[width=\columnwidth]{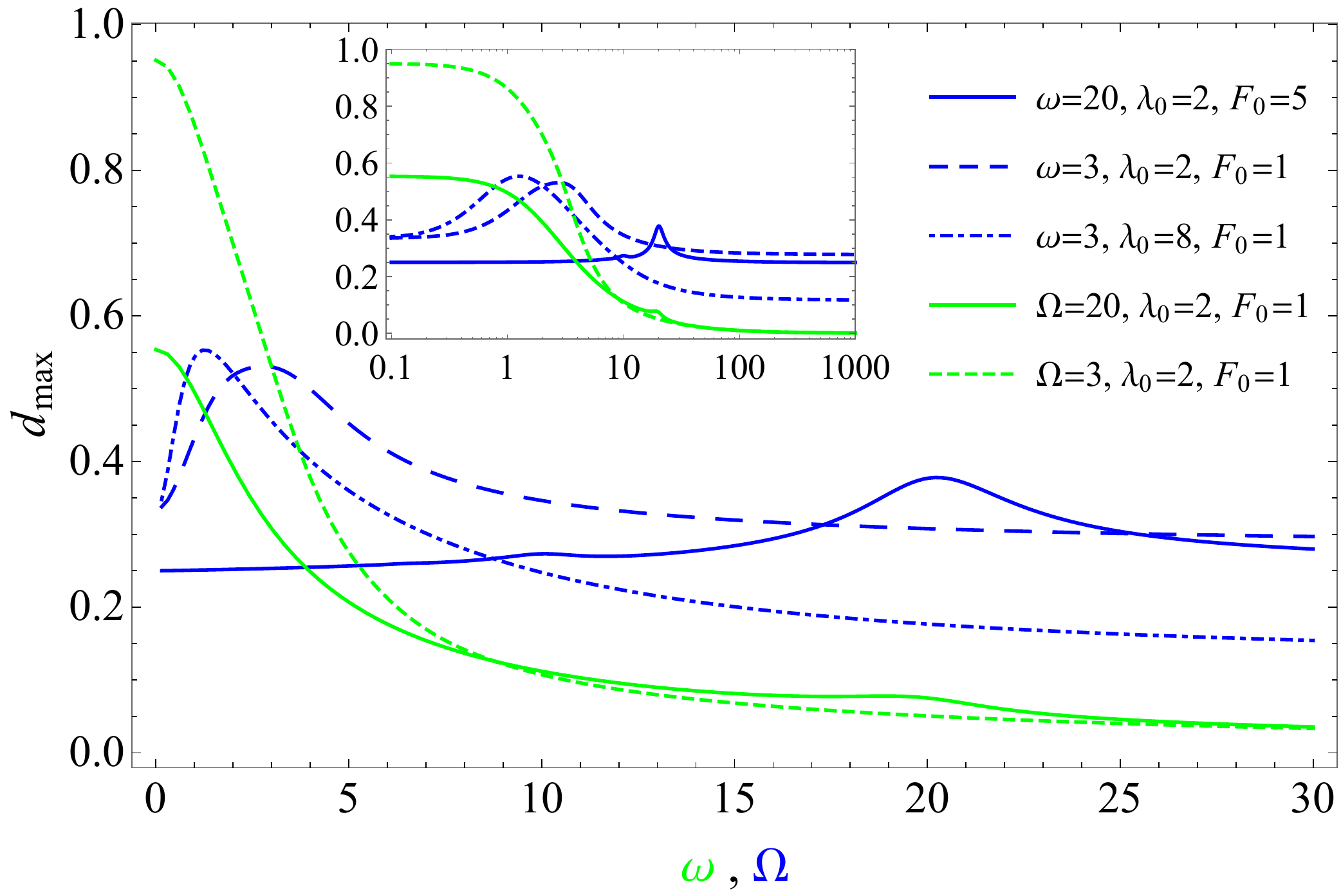}
\caption{\label{Flo:8}(Color online) Maximum distance $d_\mathrm{max}$ from the trap center of a noise-free circle swimmer with constant self-propulsion in a temporally varying harmonic trap. While the blue curves are plotted as a function of the breathing frequency $\Omega$, the green curves show the dependence on the circling frequency $\omega$. The parameters for each curve are specified in the figure. Any effect from the initial behavior of the particle has been neglected. The maximum value of $d_\mathrm{max}$ occurs at $\Omega=1.2$ for the dot-dashed blue curve, at $\Omega=2.7$ for the dashed blue curve, and at $\Omega=20.2$ for the solid blue curve. For the green curves, the maximum is always at $\omega=0$. In the inset, the asymptotic behavior for large frequencies is shown.}
\end{figure}

In analogy to the situation in Sec.\ \ref{propulsionA}, trajectories without any rotational symmetry around the trap center occur for $\omega=n\Omega$ with $n=1, 2, 3, ...$. As shown in Fig.\ \ref{sec4}(b), the condition $\Omega=\omega$ leads to a slightly distorted ellipse-like trajectory, where the center of the trap is located near a strongly bent side of the curve. 
For higher values of $n=\omega/\Omega$, more complicated swimming paths with $n-1$ inner loops are created [cf.\ Fig.\ \ref{sec4}(d)]. The generic case, however, which is realized in all situations where $\omega/\Omega$ is a noninteger number, is trajectories with an $m$-fold rotational symmetry ($m\geq 2$) around the trap center. An example of a trajectory with large $m$ is shown in Fig.\ \ref{sec4}(c).

With regard to the maximum distance $d_\mathrm{max}$ from the trap center which a particle can reach during its periodic motion, we again find a resonance situation determined by the frequencies in the system. As shown by the blue curves in Fig.\ \ref{Flo:8}, for fixed values of $\omega$ and $\lambda_{0}$ there is a clear maximum of  $d_\mathrm{max}$ as a function of $\Omega$. While the position of this maximum is usually close to $\Omega=\omega$, for large values of the trap strength $\lambda_0$ it is shifted toward lower breathing frequencies $\Omega$. This is illustrated by the dot-dashed blue curve in Fig.\ \ref{Flo:8} as compared with the dashed blue curve. Interestingly, the maximum value of $d_\mathrm{max}$ that the particle can reach is greater in the former case. This is counterintuitive at first sight because the only difference between the two cases is the trap strength, which is higher for the dot-dashed curve. Thus, one would expect that the circle swimmer is more strongly confined in the region close to the trap center. However, since the maximum occurs at a smaller breathing frequency, the time interval during which the temporally oscillating potential is weak so that the particle can escape from the trap center is extended. This overcompensates the increased trap strength. 

As illustrated by the green curves in Fig.\ \ref{Flo:8}, the maximum value of $d_\mathrm{max}$ as a function of the circling frequency $\omega$ for fixed $\Omega$ and $\lambda_0$ always occurs at $\omega=0$.  This is plausible since in that case the particle orientation is constant so that the full swimming strength can be used in order to move straight away from the trap center. For very large values of $\omega$, $d_\mathrm{max}$ goes to zero. 

Another feature of the maximum escape distance of a circle swimmer with constant self-propulsion in a temporally varying harmonic trap is the appearance of a second local maximum in addition to the global maximum. As a function of $\Omega$, this local maximum gets more pronounced when increasing the ratio $\omega/\lambda_{0}$. It occurs approximately at half the value of the breathing frequency at the global maximum (cf.\ solid blue curve in Fig.\ \ref{Flo:8}). When $d_\mathrm{max}$ is plotted as a function of the circling frequency $\omega$, a small local maximum becomes visible at $\omega \approx \Omega$ for high ratios $\Omega/\lambda_{0}$ (cf.\ solid green curve in Fig.\ \ref{Flo:8}).

\subsection{Effect of Brownian noise}

\begin{figure}
\centering
\includegraphics[width=\columnwidth]{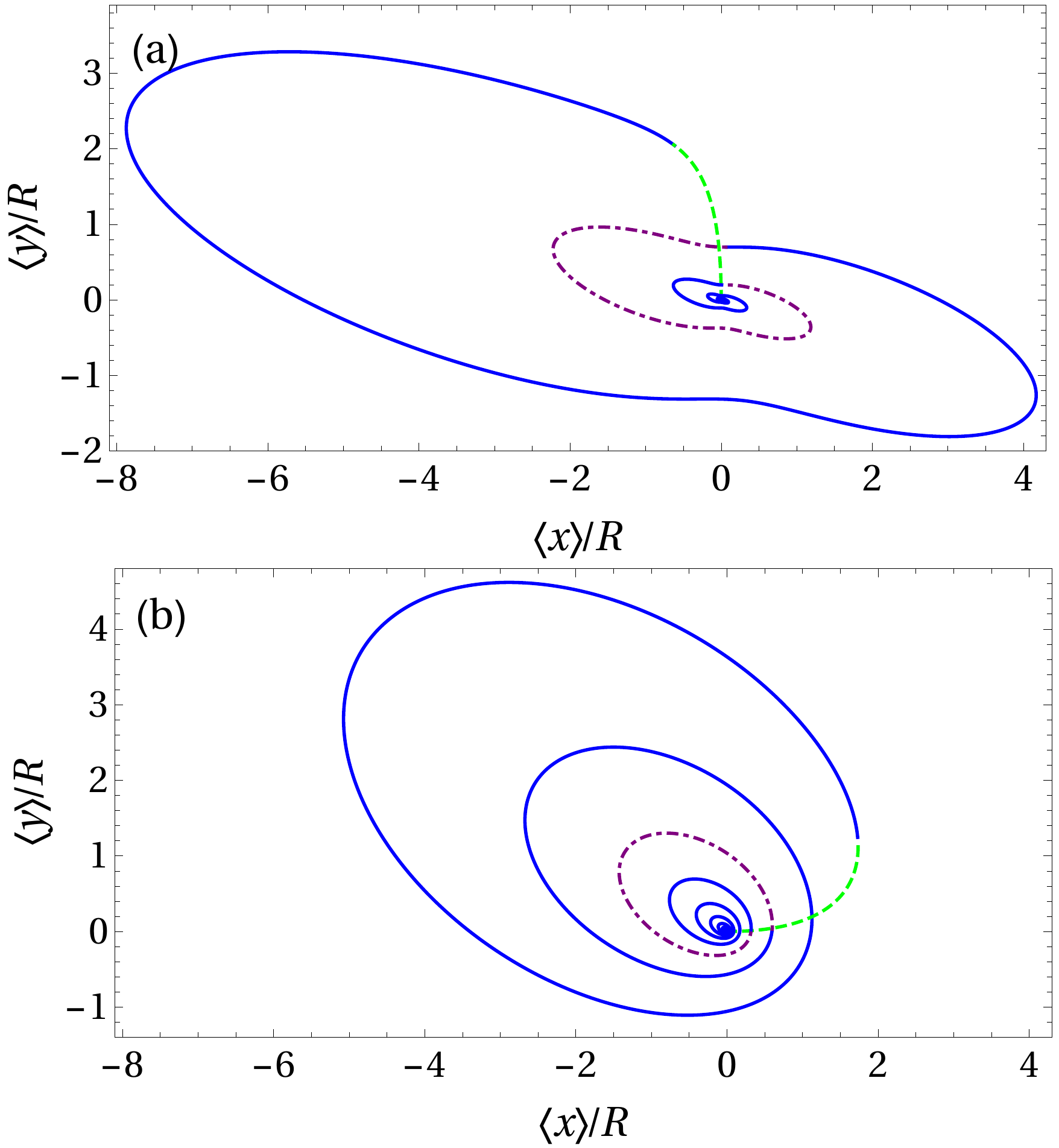}
\caption{\label{fig14}Noise-averaged trajectories of a circle swimmer with constant self-propulsion in a temporally varying harmonic trap. Both plots are for $F_{0}=50$, $\lambda_{0}=10$, and $\Omega=10$ while the values of the other parameters are $\omega=5$ and $\varphi_{0}=\pi/2$ in (a) and $\omega=10$ and $\varphi_{0}=0$ in (b). In each plot, the initial regime is visualized by the dashed green curve, and the dot-dashed purple segment represents the mean swimming path during one period.}
\end{figure}

Concerning the influence of Brownian noise, most of the findings for the case of a circle swimmer with time-dependent self-propulsion in a constant trap can analogously be transferred to the situation of a circle swimmer with constant self-propulsion in a temporally varying trap. Similar to Sec.\ \ref{propulsionB}, the main effect of the thermal noise is that self-similar mean trajectories are established (see Fig.\ \ref{fig14}), which bear the characteristics of their noise-free counterparts. Subsequent to an initial regime, the curves collapse into the trap center with the period $T_{p}=\frac{2\pi}{D_{r}}\,\mathrm{LCM}\left(\frac{1}{\omega},\frac{1}{\Omega}\right)$ of the self-similar pattern, exactly like the period of the same situation in the noise-free case. In Fig.\ \ref{fig14}, the mean swimming path during one period $T_{p}$ is visualized by the dot-dashed purple curve. Segments of the same shape but with decreasing size are continuously repeated as time proceeds. 
When the circling frequency and the breathing frequency
are equal, the self-similar trajectory is a distorted \textit{spira mirabilis}, which surrounds the trap center [cf.\ Fig.\ \ref{fig14}(b)].

\section{Experimental realization}
\label{experiment}
There are various experimental setups representing the situation described by our theoretical model. Active Brownian particles with a
programmed propulsion velocity in an external field can in principle
be realized at wish by using self-diffusiophoretic colloidal Janus particles
\cite{Volpe2011,Buttinoni2012,Kummel2013} where
both the self-propulsion and the time-dependent confinement can be
controlled independently by using laser fields with different wavelengths. The active rotational motion can be implemented either by using asymmetric particles \cite{Kummel2013} or---for spherical Janus particles---by specifically designed inhomogeneities in the gold or platinum cap \cite{Archer2015}.  
 One important aspect when preparing the experimental setup is to thoroughly take care of the optical properties of the Janus particles. Since the intensities required for optical tweezers are relatively high, the interactions of a two-component Janus particle with the light field are in general more complicated than for simple homogeneous colloidal particles. This might even lead to situations where it is not possible to trap Janus particles by using setups which perfectly work for passive colloidal particles. Therefore, the optical land\-scape has to be designed specifically for the system under study.

An alternative option is to use acoustic tweezers \cite{Shi2009} instead of optical landscapes for the trapping. This method has recently been applied successfully to self-propelled Janus particles driven by the catalytic decomposition of hydrogen peroxide \cite{Takatori2016}. An important advantage of this acoustic confinement is the possibility of designing a near-harmonic trap with a trapping radius being significantly larger than the size of the swimmer.

Finally, another experimental situation corresponding to our theoretical model is active granular hoppers on a vibrating plate. Such systems have first been realized by using self-propelled polar granular rods \cite{Narayan2007,Kudrolli2008}. In that case the direction of the active motion is determined by the particle shape. Similarly, it is also possible to include an active rotational motion by designing appropriate asymmetric particle shapes. However, a deterministic torque can also be achieved for isotropic particles by implementing the asymmetry not in the shape of the particles themselves but in the design of their legs, which are in contact with the vibrating plate \cite{Deseigne2010,Deseigne2012}. In such a system, the translational and rotational activity can be tuned by the amplitude and frequency of the vibrations. The confining potential can be realized by a corrugated surface.

All of these experiments are feasible in principle but require a thorough preparation and execution.

\section{Conclusions and outlook}
\label{conclusions}

In conclusion, we have investigated the dynamics of a Brownian circle swimmer in an external harmonic potential. By including a time dependence both with regard to the self-propulsion velocity and for the spatial potential, we have found an interesting interplay of different frequencies.  

In the absence of thermal noise, periodic trajectories are observed after passing an initial regime. The period of the trajectories is governed by the frequencies characterizing the circle swimming, the time-dependent self-propulsion, and the oscillating external potential, respectively. The existence of the initial regime is a direct consequence of the presence of the external potential. If thermal fluctuations are included, the mean trajectories for fixed initial orientation are spiral curves collapsing toward the trap center. They show a characteristic self-similar behavior for long times. The period of the self-similarity is determined by the common period of the frequencies of the system.

Moreover, the maximum spatial range a particle can explore inside the trap has been investigated. As a function of the trap strength and the various frequencies of the system, there is a resonance situation in which the swimmer can escape the maximum distance from the trap center. Changing the strength of the self-propulsion does not affect the resonance frequency but only alters the value of the maximum possible distance.

We have also provided a general result for the mean square displacement, which we explicitly solved analytically for the case of a time-dependent self-propulsion and a constant external potential. On top of a detailed analysis of the influence of the propulsion frequency, we have addressed several limiting cases. In general, for long times, the mean square displacement oscillates around a constant mean value. 

For the future, it is interesting to investigate the dynamics of circle swimmers in more complicated external potentials \cite{Bewerunge2016,Evers2013}. While a single harmonic trap is the simplest example of spatial confinement, a versatile trapping and escaping behavior is expected for periodic structures such as a staircase-like potential or traveling waves \cite{Geiseler2016PRE}. An intricate interplay between the different frequencies of the system will determine whether a particle is trapped in the first valley of a static staircase potential or is able to proceed to subsequent valleys in order to approach the global minimum of the potential. 
In this context, a consequential next step would be to generalize previous results on the escape rate of self-propelled particles from a metastable potential well \cite{Geiseler2016EPJB} toward circle swimmers.

Another interesting topic is the collective behavior of circle swimmers in a harmonic trap. It is expected that the formation of clusters as has been observed in the context of passive particles \cite{Bubeck1999,Melzer2001,Arp2004,Bonitz2006,Apolinario2006,Bonitz2010} is strongly affected by the chiral motion of active circle swimmers. In such systems of higher particle density, hydrodynamic particle-particle interactions will also play an important role and thus have to be included in the theoretical modeling \cite{Hennes2014}.

Last but not least, the investigation of chiral microswimmers in various kinds of external confinement will most likely provide valuable insight for the design of functional micromachines that have the ability to transport nanoscale cargoes in a liquid environment \cite{Baraban2012SM,Palacci2013JACS}, for example. 
These devices can even exploit various kinds of taxes \cite{Hong2007,Campbell2013,tenHagen2014NatComm,Lozano2016}, which in combination with confinement effects \cite{DiLuzio2005,vanTeeffelen2009,Fily2014,Mathijssen2016} helps them to navigate through complex environments.
Self-propelled particles combined with temporally varying harmonic potentials have recently also been suggested as a key building block for the realization of micrometer-sized heat engines \cite{Schmiedl2008,Tu2014,Krishnamurthy2016}.
To optimize the efficiency of such stochastic heat engines, a detailed understanding of the underlying processes and the competition between translational and rotational motion as characteristic for circle swimming is an important prerequisite.

\acknowledgments
This work was supported by the Deutsche Forschungsgemeinschaft (DFG, German Research Foundation) through the priority
program SPP 1726 on microswimmers under Contract No.\ LO 418/17-1 and by the European Research Council (ERC) Advanced Grant INTERCOCOS
(Grant No. 267499). B.t.H.\ gratefully acknowledges financial support through a Postdoctoral
Research Fellowship from the Deutsche Forschungsgemeinschaft -- HA 8020/1-1. 

\appendix
\section{General results for a circle swimmer in a harmonic trap}
\label{appendix}

In the following, we present the results for the most general case of a circle swimmer with time-dependent self-propulsion in a temporally varying harmonic trap. The given theoretical framework also contains the situations discussed before in detail as special cases. The general equations of motion in a nondimensionless form are
\begin{equation}
\frac{d}{dt}x\left(t\right)=\beta D\left[F\left(t\right)\cos\left(\varphi(t)\right)-\lambda\left(t\right)x\left(t\right)\right]+\sqrt{2D}\,\xi_{x}(t)
\end{equation}
and
\begin{equation}
\frac{d}{dt}y\left(t\right)=\beta D\left[F\left(t\right)\sin\left(\varphi(t)\right)-\lambda\left(t\right)y\left(t\right)\right]+\sqrt{2D}\,\xi_{y}(t).
\end{equation} 
In the absence of random forces and torques, after an initial regime the circle swimmer moves on a periodic trajectory with the period $T_{p}=\frac{2\pi}{D_{r}}\,\mathrm{LCM}\left(\frac{1}{\omega},\frac{1}{\Omega},\frac{1}{\nu}\right)$.
By averaging over the Brownian noise, the mean position
along the $x$ direction is obtained as
\begin{alignat}{1}
&\frac{\left\langle x\left(t\right)\right\rangle }{R} =c_{x}\,\exp\left[-\frac{\lambda_{0}}{\Omega}\,\sin\left(\Omega D_{r}t\right)-\lambda_{0} D_{r}t\right]\nonumber\\&\quad+\left\{ F{}_{0}\exp\left[-\frac{\lambda_{0}}{\Omega}\,\sin\left(\Omega D_{r}t\right)-\lambda_{0} D_{r}t\right]\right.\nonumber \\
 &\quad\times\int_{0}^{D_{r}t}d\tau\:\cos\big(\omega \tau+\varphi_{0}\big)\Big(1+\cos\left(\nu\tau+\theta\right)\Big)\nonumber \\
 &\quad\quad\quad \left.\times\exp\left[-\tau+\frac{\lambda_{0}}{\Omega}\,\sin\left(\Omega\tau\right)+\lambda_{0} \tau\right]\right\}.
\label{appx}
\end{alignat}
Correspondingly, the $y$ component is 
\begin{alignat}{1}
&\frac{\left\langle y\left(t\right)\right\rangle }{R} =c_{y}\,\exp\left[-\frac{\lambda_{0}}{\Omega}\,\sin\left(\Omega D_{r}t\right)-\lambda_{0} D_{r}t\right]\nonumber\\&\quad+\left\{ F{}_{0}\exp\left[-\frac{\lambda_{0}}{\Omega}\,\sin\left(\Omega D_{r}t\right)-\lambda_{0} D_{r}t\right]\right.\nonumber \\
 &\quad\times\int_{0}^{D_{r}t}d\tau\:\sin\big(\omega \tau+\varphi_{0}\big)\Big(1+\cos\left(\nu\tau+\theta\right)\Big)\nonumber \\
 &\quad\quad\quad \left.\times\exp\left[-\tau+\frac{\lambda_{0}}{\Omega}\,\sin\left(\Omega\tau\right)+\lambda_{0} \tau\right]\right\}.
\label{appy}
\end{alignat}
Equations \eqref{appx} and \eqref{appy} represent self-similar noise-averaged swimming paths, which collapse into the trap center.  

The second moment is given by $\left\langle r^{2}\left(t\right)\right\rangle /R^{2}=\left\langle x^{2}\left(t\right)\right\rangle /R^{2}+\left\langle y^{2}\left(t\right)\right\rangle /R^{2}$
with
\begin{widetext}
\begin{alignat}{1}
&\frac{\left\langle x^{2}\left(t\right)\right\rangle }{R^{2}} =\exp\left(-2\frac{\lambda_{0}}{\Omega}\,\sin\left(\Omega D_{r}t\right)-2\lambda{}_{0}D_{r}t\right) c_{x}^{2} +\exp\left(-2\frac{\lambda_{0}}{\Omega}\,\sin\left(\Omega D_{r}t\right)-2\lambda{}_{0}D_{r}t\right)\nonumber\\ & \times\Biggl\{\biggl[2c_{x}F{}_{0}\int_{0}^{D_{r}t}d\tau\:\cos\left(\omega \tau+\varphi_{0}\right)\Big(1+\cos\left(\nu\tau+\theta\right)\Big)\times\exp\left(-\tau+\frac{\lambda{}_{0}}{\Omega}\,\sin\left(\Omega\tau\right)+\lambda_{0}\tau\right)\biggr]\nonumber \\
 & +\int_{0}^{D_{r}t}d\tau_{1}\int_{0}^{\tau_{1}}d\tau_{2}\biggl[F_{0}^{2}\exp\left[-\left(\tau_{1}-\tau_{2}\right)\right] \left(1+\cos\left(\nu\tau_{1}+\theta\right)\right)\left( 1+\cos\left(\nu\tau_{2}+\theta\right)\right)\nonumber \\
 & \times\bigg(\cos\Big(\omega\big(\tau_{1}- \tau_{2}\big)\Big)+\cos\Big(2\varphi_{0}+\omega\big(\tau_{1}+\tau_{2}\big)\Big) e^{-4\tau_{2}}\bigg)\exp\left(\frac{\lambda{}_{0}}{\Omega}\,\sin\left(\Omega\tau_{1}\right)+\lambda_{0}\tau_{1}\right)\nonumber \\
 & \times\exp\left(\frac{\lambda{}_{0}}{\Omega}\,\sin\left(\Omega\tau_{2}\right)+\lambda_{0}\tau_{2}\right)\biggr] +\frac{8}{3}\int_{0}^{D_{r}t}d\tau\exp\left(2\frac{\lambda{}_{0}}{\Omega}\,\sin\left(\Omega\tau\right)+2\lambda_{0}\tau\right)\Biggr\}
\end{alignat}
and
\begin{alignat}{1}
&\frac{\left\langle y^{2}\left(t\right)\right\rangle }{R^{2}} =\exp\left(-2\frac{\lambda_{0}}{\Omega}\,\sin\left(\Omega D_{r}t\right)-2\lambda{}_{0}D_{r}t\right) c_{y}^{2} +\exp\left(-2\frac{\lambda_{0}}{\Omega}\,\sin\left(\Omega D_{r}t\right)-2\lambda{}_{0}D_{r}t\right)\nonumber\\ & \times\Biggl\{\biggl[2c_{y}F{}_{0}\int_{0}^{D_{r}t}d\tau\:\sin\left(\omega \tau+\varphi_{0}\right)\Big(1+\cos\left(\nu\tau+\theta\right)\Big)\exp\left(-\tau+\frac{\lambda{}_{0}}{\Omega}\,\sin\left(\Omega\tau\right)+\lambda_{0}\tau\right)\biggr]\nonumber \\
 & +\int_{0}^{D_{r}t}d\tau_{1}\int_{0}^{\tau_{1}}d\tau_{2}\biggl[F_{0}^{2}\exp\left(-\left(\tau_{1}-\tau_{2}\right)\right)\left(1+\cos\left(\nu\tau_{1}+\theta\right)\right)\left( 1+\cos\left(\nu\tau_{2}+\theta\right)\right)\nonumber \\
 & \times\bigg(\cos\Big(\omega\big(\tau_{1}- \tau_{2}\big)\Big)-\cos\Big(2\varphi_{0}+\omega\big(\tau_{1}+\tau_{2}\big)\Big) e^{-4\tau_{2}}\bigg)\exp\left(\frac{\lambda{}_{0}}{\Omega}\,\sin\left(\Omega\tau_{1}\right)+\lambda_{0}\tau_{1}\right)\nonumber \\
 & \times\exp\left(\frac{\lambda{}_{0}}{\Omega}\,\sin\left(\Omega\tau_{2}\right)+\lambda_{0}\tau_{2}\right)\biggr] +\frac{8}{3}\int_{0}^{D_{r}t}d\tau\exp\left(2\frac{\lambda{}_{0}}{\Omega}\,\sin\left(\Omega\tau\right)+2\lambda_{0}\tau\right)\Biggr\}.
\end{alignat}
These equations can be solved analytically for a constant spatial trap. In that case, by defining
\begin{align}
 G_{1} & =\left(\lambda_{0}+1\right)\left\{F_{0} \frac{e^{2\lambda_{0}D_{r}t}-1}{2\lambda_{0}}+f\left(D_{r}t,\nu,\theta,2\lambda_{0}\right)-f\left(D_{r}t,\omega,0,\lambda_{0}-1\right)-f_{c}\left(D_{r}t,\omega,0,\nu,\theta,\lambda_{0}-1\right)\right\}\nonumber \\
 & +\omega\left\{ f\left(D_{r}t,\omega,-\frac{\pi}{2},\lambda_{0}-1\right)+f_{c}\left(D_{r}t,\nu,\theta,\omega,-\frac{\pi}{2},\lambda_{0}-1\right)\right\},
  \end{align}
\begin{alignat}{1}
G_{2} &=\left(\lambda_{0}+1\right)\left\{ f\left(D_{r}t,\nu,\theta,2\lambda_{0}\right)+f_{c}\left(D_{r}t,\nu,\theta,\nu,\theta,2\lambda_{0}\right)\right\} -\left(\omega-\nu\right)\Big\{ f\left(D_{r}t,\nu,\theta-\frac{\pi}{2},2\lambda_{0}\right)\nonumber \\
 & +f_{c}\left(D_{r}t,\nu^{'},\theta,\nu,\theta-\frac{\pi}{2},2\lambda_{0}\right)\Big\} -\left(\lambda_{0}+1\right)\Big\{ f\left(D_{r}t,\omega,\theta,\lambda_{0}-1\right)+f_{c}\left(D_{r}t,\nu,\theta,\omega,\theta,\lambda_{0}-1\right)\Big\}\nonumber \\
 & +\left(\omega-\nu\right)\left\{ f\left(D_{r}t,\omega,\theta-\frac{\pi}{2},\lambda_{0}-1\right)+f_{c}\left(D_{r}t,\nu,\theta,\omega,\theta-\frac{\pi}{2},\lambda_{0}-1\right)\right\}, 
 \end{alignat}
and
\begin{align}
G_{3} &=\left(\lambda_{0}+1\right)\left\{ f\left(D_{r}t,\nu,\theta,2\lambda_{0}\right)+f_{c}\left(D_{r}t,\nu,\theta,\nu,\theta,2\lambda_{0}\right)\right\} +\left(\omega+\nu\right)\Big\{ f\left(D_{r}t,\nu,\theta-\frac{\pi}{2},2\lambda_{0}\right)\nonumber \\
 &+f_{c}\left(D_{r}t,\nu,\theta,\nu,\theta-\frac{\pi}{2},2\lambda_{0}\right)\Big\}-\left(\lambda_{0}+1\right)\left\{ f\left(D_{r}t,\omega,\theta,\lambda_{0}-1\right)+f_{c}\left(D_{r}t,\nu,\theta,\omega,\theta,\lambda_{0}-1\right)\right\}\nonumber \\
 & +\left(\omega+\nu\right)\left\{ f\left(D_{r}t,\omega,\theta-\frac{\pi}{2},\lambda_{0}-1\right)+f_{c}\left(D_{r}t,\nu,\theta,\omega,\theta-\frac{\pi}{2},\lambda_{0}-1\right)\right\}
\end{align}
with $f_{c}\left(t,\nu,\theta,\omega,\varphi,\lambda\right)  =\frac{1}{2}f\left(t,\omega+\nu,\varphi+\theta,\lambda\right)+\frac{1}{2}f\left(t,\omega-\nu,\varphi-\theta,\lambda\right)$, one obtains
\begin{alignat}{1}
\frac{\left\langle x^{2}(t)\right\rangle +\left\langle y^{2}(t)\right\rangle }{R^{2}} & =2e^{-\lambda_{0}D_{r}t}\left(c_{x}\frac{\left\langle x\left(t\right)\right\rangle }{R}+c_{y}\frac{\left\langle y\left(t\right)\right\rangle }{R}\right) -\left(c_{x}^{2}+c_{y}^{2}\right)e^{-2\lambda_{0}D_{r}t}+ \frac{8}{3\lambda_{0}}\left(1-e^{-2\lambda_{0}D_{r}t}\right)\nonumber \\
 & +F_{0}e^{-2\lambda_{0}D_{r}t}\left(\frac{2G_{1}}{\left(\lambda_{0}+1\right)^{2}+\omega^{2}}+\frac{G_{2}}{\left(\lambda_{0}+1\right)^{2}+\left(\omega-\nu\right)^{2}}+\frac{G_{3}}{\left(\lambda_{0}+1\right)^{2}+\left(\omega+\nu\right)^{2}}\right).
\label{eq:msd}
 \end{alignat}
 In the above formulas, the function $f$ is defined by Eq.\ \eqref{II3} and the mean positions $\left\langle x\left(t\right)\right\rangle/R$ and $\left\langle y\left(t\right)\right\rangle/R$ are given in Eqs.\ \eqref{IIB1} and \eqref{IIB2}, respectively. For $c_x=c_y=0$, the second moment in Eq.\ \eqref{eq:msd} is identical to the mean square displacement visualized in Fig.\ \ref{msd}(b).

\end{widetext}

\bibliography{refs}

\end{document}